
\input harvmac.tex
\input amssym.tex 
\overfullrule=0pt 



    \input epsf
    
    \def\figscale#1#2{\epsfxsize=#2\epsfbox{#1.eps}}
\def\caption#1#2{{\ninepoint\noindent Fig~#1. #2}\medskip}





\def\a{\alpha}\def\b{\beta}\def\d{\delta}\def\e{\epsilon}
\def\g{\gamma}\def\k{\kappa}\def\l{\lambda}\def\m{\mu}\def\n{\nu}
\def\o{\omega}\def\r{\rho}\def\s{\sigma}\def\t{\tau}\def\z{\zeta}
\def\D{\Delta}\def\G{\Gamma}\def\L{\Lambda}\def\O{\Omega}

\def\CF{{\cal F}}\def\CI{{\cal I}}\def\CM{{\cal M}}
\def\CN{{\cal N}}\def\CV{{\cal V}}

\def\IC{{\Bbb C}} 
\def\IP{{\Bbb P}} 
\def\IR{{\Bbb R}} 
\def\IZ{{\Bbb Z}} 
\def\IF{{\Bbb F}} 
\def\so{{\frak so}}

\def\det{\mathop{\rm det}\nolimits}

\def\Re{\mathop{\rm Re}\nolimits}
\def\Im{\mathop{\rm Im}\nolimits}
\def\Vol{\mathop{\rm Vol}\nolimits}


\def\tilde{\widetilde}

\def\w{\wedge}
\def\ha{{1\over2}}
\def\half{{\textstyle{1\over2}}}
\def\quarter{{\textstyle{1\over4}}}

\def\iover#1{{\textstyle{i\over{#1}}}}
\def\smallfrac#1#2{{\textstyle{{#1}\over{#2}}}}

\def\free{{\rm free}}
\def\tor{{\rm tor}}
\def\NS{{\rm NS}}
\def\RR{{\rm RR}}
\def\TZZ{{T^6/(\IZ_2\times \IZ_2)}}
\def\footlabel#1{\xdef#1{\the\ftno}} 
\def\nonnumberedsubsec#1{\ifnum\lastpenalty>9000\else\bigbreak\fi
\noindent{\it{#1}}\par\nobreak\medskip\nobreak}



\lref\AldazabalUP{
  G.~Aldazabal, P.~G.~C\'amara, A.~Font and L.~E.~Ib\'a\~nez,
  ``More dual fluxes and moduli fixing,''
  JHEP {\bf 0605}, 070 (2006)
  [arXiv:hep-th/0602089].
}
\lref\AspinwallMN{
  P.~S.~Aspinwall,
  ``K3 surfaces and string duality,''
  arXiv:hep-th/9611137.
}
\lref\BeckerSH{
  K.~Becker, M.~Becker, P.~S.~Green, K.~Dasgupta and E.~Sharpe,
  ``Compactifications of heterotic strings on non-K\"ahler complex manifolds.
  II,''
  Nucl.\ Phys.\ B {\bf 678}, 19 (2004)
  [arXiv:hep-th/0310058].
}
\lref\BehrndtMJ{
  K.~Behrndt and M.~Cveti\v c,
  ``General $\CN = 1$ supersymmetric fluxes in massive type IIA string
  theory,''
  Nucl.\ Phys.\ B {\bf 708}, 45 (2005)
  [arXiv:hep-th/0407263].
}
\lref\BenmachicheDF{
  I.~Benmachiche and T.~W.~Grimm,
  ``Generalized $\CN = 1$ orientifold compactifications and the Hitchin
  functionals,''
  Nucl.\ Phys.\  B {\bf 748}, 200 (2006)
  [arXiv:hep-th/0602241].
}
\lref\BlumenhagenCI{
  R.~Blumenhagen, B.~K\"ors, D.~L\"ust and S.~Stieberger,
  ``Four-dimensional string compactifications with D-branes,
  orientifolds and fluxes,'' 
  arXiv:hep-th/0610327.
}
\lref\BlumenhagenXT{
  R.~Blumenhagen, M.~Cveti\v c and T.~Weigand,
  ``Spacetime instanton corrections in 4D string vacua---the seesaw mechanism
  for D-brane models,''
  arXiv:hep-th/0609191.
}
\lref\BlumenhagenXX{
  R.~Blumenhagen, F.~Gmeiner, G.~Honecker, D.~L\"ust and T.~Weigand,
  ``The statistics of supersymmetric D-brane models,''
  Nucl.\ Phys.\ B {\bf 713}, 83 (2005)
  [arXiv:hep-th/0411173].
}
\lref\CamaraDC{
  P.~G.~C\'amara, A.~Font and L.~E.~Ib\'a\~nez,
  ``Fluxes, moduli fixing and MSSM-like vacua in a simple IIA orientifold,''
  JHEP {\bf 0509}, 013 (2005)
  [arXiv:hep-th/0506066].
}
\lref\CveticAC{
  M.~Cveti\v c,
  ``Blown-Up Orbifolds,''
  SLAC-PUB-4325,
  {\it Invited talk given at Int. Workshop on Superstrings, Composite
  Structures, and Cosmology, College Park, MD, Mar 11-18, 1987}
}
\lref\CveticTJ{
  M.~Cveti\v c, G.~Shiu and A.~M.~Uranga,
  ``Three-family supersymmetric standard like models from intersecting brane
  worlds,''
  Phys.\ Rev.\ Lett.\  {\bf 87}, 201801 (2001)
  [arXiv:hep-th/0107143].
}
\lref\DabholkarSY{
  A.~Dabholkar and C.~Hull,
  ``Duality twists, orbifolds, and fluxes,''
  JHEP {\bf 0309}, 054 (2003)
  [arXiv:hep-th/0210209].
}
\lref\DabholkarVE{
  A.~Dabholkar and C.~Hull,
  ``Generalised T-duality and non-geometric backgrounds,''
  JHEP {\bf 0605}, 009 (2006)
  [arXiv:hep-th/0512005].
}
\lref\DasguptaSS{
  K.~Dasgupta, G.~Rajesh and S.~Sethi,
  ``M theory, orientifolds and $G$-flux,''
  JHEP {\bf 9908}, 023 (1999)
  [arXiv:hep-th/9908088].
}
\lref\DenefMM{
  F.~Denef, M.~R.~Douglas, B.~Florea, A.~Grassi and S.~Kachru,
  ``Fixing all moduli in a simple F-theory compactification,''
  arXiv:hep-th/0503124.
}
\lref\DerendingerPH{
  J.~P.~Derendinger, C.~Kounnas, P.~M.~Petropoulos and F.~Zwirner,
  ``Fluxes and gaugings: $\CN = 1$ effective superpotentials,''
  Fortsch.\ Phys.\  {\bf 53}, 926 (2005)
  [arXiv:hep-th/0503229].
}
\lref\DouglasES{
  M.~R.~Douglas and S.~Kachru,
  ``Flux compactification,''
  arXiv:hep-th/0610102.
}
\lref\FerraraYX{
  S.~Ferrara, J.~A.~Harvey, A.~Strominger and C.~Vafa,
  ``Second Quantized Mirror Symmetry,''
  Phys.\ Lett.\ B {\bf 361}, 59 (1995)
  [arXiv:hep-th/9505162].
}
\lref\GoldsteinPG{
  E.~Goldstein and S.~Prokushkin,
  ``Geometric model for complex non-K\"ahler manifolds with $SU(3)$
  structure,'' 
  Commun.\ Math.\ Phys.\  {\bf 251}, 65 (2004)
  [arXiv:hep-th/0212307].
}
\lref\GranaBG{
  M.~Grana, R.~Minasian, M.~Petrini and A.~Tomasiello,
  ``Supersymmetric backgrounds from generalized Calabi-Yau manifolds,''
  JHEP {\bf 0408}, 046 (2004)
  [arXiv:hep-th/0406137].
}
\lref\GranaSN{
  M.~Gra\~na, R.~Minasian, M.~Petrini and A.~Tomasiello,
  ``Generalized structures of $\CN = 1$ vacua,''
  JHEP {\bf 0511}, 020 (2005)
  [arXiv:hep-th/0505212].
}
\lref\GrimmTM{
  T.~W.~Grimm, A.~Klemm, M.~Marino and M.~Weiss,
  ``Direct integration of the topological string,''
  arXiv:hep-th/0702187.
}
\lref\GrimmUA{
  T.~W.~Grimm and J.~Louis,
  ``The effective action of type IIA Calabi-Yau orientifolds,''
  Nucl.\ Phys.\ B {\bf 718}, 153 (2005)
  [arXiv:hep-th/0412277].
}
\lref\GrossHC{
  M.~Gross,
  ``Topological mirror symmetry,''
  arXiv:math.ag/9909015.
}
\lref\HeadrickCH{
  M.~Headrick and T.~Wiseman,
  ``Numerical Ricci-flat metrics on $K3$,''
  Class.\ Quant.\ Grav.\  {\bf 22}, 4931 (2005)
  [arXiv:hep-th/0506129].
}
\lref\HellermanAX{
  S.~Hellerman, J.~McGreevy and B.~Williams,
  ``Geometric constructions of nongeometric string theories,''
  JHEP {\bf 0401}, 024 (2004)
  [arXiv:hep-th/0208174].
}

\lref\HellermanTX{
  S.~Hellerman and J.~Walcher,
  ``Worldsheet CFTs for flat monodrofolds,''
  arXiv:hep-th/0604191.
}
\lref\HullHK{
  C.~M.~Hull and R.~A.~Reid-Edwards,
  ``Flux compactifications of string theory on twisted tori,''
  arXiv:hep-th/0503114.
}
\lref\HullIN{
  C.~M.~Hull,
  ``A geometry for non-geometric string backgrounds,''
  JHEP {\bf 0510}, 065 (2005)
  [arXiv:hep-th/0406102].
}
\lref\HullTP{
  C.~M.~Hull and R.~A.~Reid-Edwards,
  ``Flux compactifications of M-theory on twisted tori,''
  JHEP {\bf 0610}, 086 (2006)
  [arXiv:hep-th/0603094].
}
\lref\KachruAW{
  S.~Kachru, R.~Kallosh, A.~Linde and S.~P.~Trivedi,
  ``De Sitter vacua in string theory,''
  Phys.\ Rev.\ D {\bf 68}, 046005 (2003)
  [arXiv:hep-th/0301240].
}
\lref\KachruHE{
  S.~Kachru, M.~B.~Schulz and S.~Trivedi,
  ``Moduli stabilization from fluxes in a simple IIB orientifold,''
  JHEP {\bf 0310}, 007 (2003)
  [arXiv:hep-th/0201028].
}
\lref\KaloperYR{
  N.~Kaloper and R.~C.~Myers,
  ``The O(dd) story of massive supergravity,''
  JHEP {\bf 9905}, 010 (1999)
  [arXiv:hep-th/9901045].
}
\lref\KasteXS{
  P.~Kaste, R.~Minasian, M.~Petrini and A.~Tomasiello,
  ``Kaluza-Klein bundles and manifolds of exceptional holonomy,''
  JHEP {\bf 0209}, 033 (2002)
  [arXiv:hep-th/0206213].
}
\lref\LawrenceMA{
  A.~Lawrence, M.~B.~Schulz and B.~Wecht,
  ``D-branes in nongeometric backgrounds,''
  JHEP {\bf 0607}, 038 (2006)
  [arXiv:hep-th/0602025].
}
\lref\LustIG{
  D.~L\"ust and D.~Tsimpis,
  ``Supersymmetric AdS$_4$ compactifications of IIA supergravity,''
  JHEP {\bf 0502}, 027 (2005)
  [arXiv:hep-th/0412250].
}
\lref\MaharanaMY{
  J.~Maharana and J.~H.~Schwarz,
  ``Noncompact symmetries in string theory,''
  Nucl.\ Phys.\ B {\bf 390}, 3 (1993)
  [arXiv:hep-th/9207016].
}
\lref\ScherkZR{
  J.~Scherk and J.~H.~Schwarz,
  ``How To Get Masses From Extra Dimensions,''
  Nucl.\ Phys.\ B {\bf 153}, 61 (1979).
}
\lref\SchulzUB{
  M.~B.~Schulz,
  ``Superstring orientifolds with torsion: O5 orientifolds of torus 
  fibrations and their massless spectra,''
  Fortsch.\ Phys.\  {\bf 52}, 963 (2004)
  [arXiv:hep-th/0406001].
}
\lref\SchulzTT{
  M.~B.~Schulz,
  ``Calabi-Yau duals of torus orientifolds,''
  JHEP {\bf 0605}, 023 (2006)
  [arXiv:hep-th/0412270].
}
\lref\TomasielloBP{
  A.~Tomasiello,
  ``Topological mirror symmetry with fluxes,''
  JHEP {\bf 0506}, 067 (2005)
  [arXiv:hep-th/0502148].
}
\lref\VilladoroCU{
  G.~Villadoro and F.~Zwirner,
  ``$\CN = 1$ effective potential from dual type-IIA D6/O6 orientifolds with
  general fluxes,''
  JHEP {\bf 0506}, 047 (2005)
  [arXiv:hep-th/0503169].
}
\lref\WendlandRY{
  K.~Wendland,
  ``Consistency of orbifold conformal field theories on $K3$,''
  Adv.\ Theor.\ Math.\ Phys.\  {\bf 5}, 429 (2002)
  [arXiv:hep-th/0010281].
}


\lref\BorceaTQ{
  C.~Borcea, ``$K3$ surfaces with involution and mirror pairs of
  Calabi-Yau manifolds,'' in B.~Greene and S.~T.~Yau (eds.), Mirror
  Symmetry II, Int.\ Press (1997) 717-743.
}
\lref\BottTu{
  R.~Bott, L.~Tu, {\it Differential Forms in Algebraic Topology},
  Springer-Verlag, New York (1982) 194.
}
\lref\CveticHet{
  M.~Cveti\v c, T.~Liu and M.~B.~Schulz, work in progress.
}
\lref\Griffiths{
  P.~Griffiths and J.~Harris, ``Principles of Algebraic Geometry,''
  Wiley-Interscience (1978).
}
\lref\Voisin{
  C.~Voisin, Ast\'erisque, {\bf 218}, Soc.\ Math. France (1993) 273.
}
\lref\Nikulin{
  V.~V.~Nikulin, ``Discrete reflection groups in Lobachevsky spaces
  and algebraic surfaces,''
  Proc.\ ICM, Berkeley, California (1986) 654--671.
}
\lref\SchulzKK{
  M.~B.~Schulz, ``Toward warped Kaluza-Klein reduction,'' to appear.
}


\Title{\vbox{\hbox{hep-th/0701204}\hbox{UPR-1167-T}\hbox{NSF-KITP-07-14}}}
{\vbox{\centerline{Twisting $K3\times T^2$ Orbifolds}}}
\centerline{\phantom{$^{a b 3}\,$}
Mirjam Cveti\v c $^a$\footnote{${}^1$}{cvetic at physics.upenn.edu},
Tao Liu $^a$\footnote{${}^2$}{liutao at physics.upenn.edu},
Michael B.~Schulz $^{a b}$\footnote{$^3$}{mbschulz at brynmawr.edu}}
\medskip
\centerline{\llap{$^a\,$}{\it Department of Physics and Astronomy,
University of Pennsylvania}}
\centerline{{\it Philadelphia, PA 19104, USA}}
\medskip
\centerline{\llap{$^b\,$}{\it Department of Physics, Bryn Mawr College}}
\centerline{{\it Bryn Mawr, PA 19010, USA}}
\vskip .3in

We construct a class of geometric twists of Calabi-Yau manifolds of
Voisin-Borcea type $(K3\times T^2)/\IZ_2$ and study the superpotential
in a type IIA orientifold based on this geometry.  The twists modify
the direct product by fibering the $K3$ over $T^2$ while preserving
the $\IZ_2$ involution.  As an important application, the
Voisin-Borcea class contains \hbox{$T^6/(\IZ_2\times\IZ_2)$}, the
usual setting for intersecting D6 brane model building.  Past work in
this context considered only those twists inherited from $T^6$, but
our work extends these twists to a subset of the blow-up modes.  Our
work naturally generalizes to arbitrary K3 fibered Calabi-Yau
manifolds and to nongeometric constructions.

\Date{\vbox{\hbox{22 January 2007}\hbox{Revised: 21 August 2007}}}


\newsec{Introduction}

In the quest for a better understanding of the space of string theory
vacua, it has become increasingly clear that in addition to
Neveu-Schwarz and Ramond-Ramond fluxes, the set of discrete data
defining a string compactification also includes a set of geometric
(and nongeometric) twists.  The geometric twists characterize the
departure of the internal manifold from special holonomy, in most
cases from a Calabi-Yau threefold.  One way to describe this departure
is in terms of $G$-structures and intrinsic torsion.  For $SU(3)$
structure, this means specifying the moduli-dependent quantities $dJ$
and $d\Omega$.  However, this description masks the discreteness of
the underlying geometric choice.  An alternative description without
this deficiency is to define the twists as a discrete deformation of
the Calabi-Yau cohomology ring (cf.~Sec.~2).  In this approach, a
construction of the resulting non Calabi-Yau manifold which realizes
the deformed cohomology has been lacking except in a few special
cases---the mirror of the quintic hypersurface in $\IP^4$
\TomasielloBP,\foot{To be fair, Ref.~\TomasielloBP\ has given a formal
description that applies not only to the the quintic but to any $T^3$
fibered Calabi-Yau 3-fold $X$.  However, to apply this formalism to a
given choice of $X$, the Strominger-Yau-Zaslow $T^3$ fibration needs
to be explicitly known for $X$.} certain $T^2$ fibrations over
$K3$,\foot{Heterotic vacua involving $T^2$ fibrations over $K3$ were
first discussed in Ref.~\DasguptaSS.  A more recent and thorough
account can be found in Ref.~\BeckerSH.  See also Ref.~\LustIG\ for
type IIA compactifications based on this geometry.}, and the untwisted
sector of the $\TZZ$ orbifold
\refs{\CamaraDC,\AldazabalUP,\VilladoroCU,\DerendingerPH}. Therefore,
one would like a more general construction.  In addition, given the
ubiquity of $\TZZ$ in intersecting brane models of low energy particle
physics, it is of particular interest to understand how blow-up modes
of this orbifold can participate in the geometric twists even this
special case.

In this article, we interpret $\TZZ$ as $(K3\times T^2)/\IZ_2$ and
show how the full $\IZ_2$ projected sector of the latter can be
included in the choice geometric twists.  From the point of view of
the $\TZZ$ orbifold, this corresponds to the untwisted sector plus one
third of the twisted sector, since 16 of the 48 blow-up modes of
$\TZZ$ are accounted for by the blow-up modes of $K3 =T^4/\IZ_2$.  Our
construction generalizes straightforwardly to other Calabi-Yau
threefolds of the Voisin-Borcea type (i.e., $(K3\times T^2)/\IZ_2$ for
other choices of $\IZ_2$ involution) \refs{\BorceaTQ,\Voisin}, and
more abstractly to arbitrary $K3$ fibered Calabi-Yau manifolds and
nongeometric constructions.

An outline of the paper is as follows:

In Sec.~2, we define geometric twists as discrete data that deform the
closure and non-exactness properties of the generators of the
Calabi-Yau integer cohomology ring.

Secs.~3 and 4 are the core of the paper.  In Sec.~3 and its associated
Apps.~A and B, we consider the orbifold $\TZZ$, its interpretation as
$(K3\times T^2)/\IZ_2$, and the discrete twist data needed to
parametrize a nontrivial $K3$ fibration over $T^2$ compatible with the
$\IZ_2$ involution.  Along the way, we give a pedagogical review of
the coset moduli space of $K3$ and its interpretation as a change of
basis between two natural bases of $H^2(K3,\IR)$.  We also discuss the
quantization of the twist data and how it differs in two qualitatively
different classes of $K3$ fibrations

In Sec.~4, we consider an $\CN=1$ orientifold of type IIA string
theory compactified on this twisted background.  After describing the
formal structure of the 4D $\CN =1$ supergravity
theory---superpotential, K\"ahler potential, and Bianchi identities
associated with D6 charge---we turn to an analysis of the vacua of the
theory, focusing on supersymmetric Minkowski vacua.  We find that
supersymmetric Minkowski vacua are of the holomorphic monopole form
discussed in Refs.~\refs{\KasteXS,\GranaBG}.  These type~IIA
orientifold vacua formally lift to compactifications of M~theory on
manifolds of $G_2$ holonomy, which could in principle be studied
directly.  However, in the past, the dual type~IIA D6/O6 perspective
has proven a fruitful setting for model building.  Enriching this
framework through the introduction of new geometric twist data
provides what we hope will be a useful addition to the model building
toolbox.  After discussing the validity of the classical supergravity
description, we present three examples of Minkowski vacua which share
the property that all blow-up moduli of $(K3\times T^2)/\IZ_2$ are
driven to zero.  The first of the examples includes the full metric
backreaction and dilaton profile in order to illustrate that these
features do not affect the supersymmetry constraints on moduli.  The
section ends with a short discussion of the supersymmetry conditions
in the more generic case of Anti de~Sitter vacua.

Secs.~5 and 6 are devoted to generalizations.  In Sec.~5 and its
associated App.~C, we describe a straightforward generalization from
$\TZZ$ to all Calabi-Yau manifolds of the Voisin-Borcea type
$(K3\times T^2)/\IZ_2$.  In fact, it was primarily notational
simplicity and the centrality of $\TZZ$ in model building that
motivated the focus on $\TZZ$ in Sec.~3 and 4.  The other
Voisin-Borcea manifolds would have served equally well.  In Sec.~6, we
describe further generalizations, first to the class of $K3$
fibrations over $\IP^1$, and then to nongeometric twists that employ
the full $\G_{4,20}$ duality group of type IIA string theory on
$K3$.

In Sec.~7, we conclude with a brief discussion of our results and a
description of possible avenues for future work.


\newsec{Geometric twists as discrete deformation of Calabi-Yau
cohomology}

Let us start with an arbitrary Calabi-Yau manifold $X$.  For
simplicity, we assume that the torsion part of the cohomology of $X$
vanishes, $H^\tor(X,\IZ) = 0$.  (Once we introduce geometric
twists this will change, and the discretely deformed manifold will
generically no longer be a Calabi-Yau.)  Let $\omega_a$ and
$\tilde\omega^a$ denote bases of $H^2(X,\IZ)$ and $H^4(X,\IZ)$
respectively, where the two bases have been chosen to satisfy
\eqn\TwoDotFour{\int\omega_a\wedge\tilde\omega^b = \delta_a{}^b.}
Likewise, we choose a symplectic decomposition of $H_3(X,\IZ)$
into $A$-cycles and $B$-cycles, and corresponding decomposition of the
cohomology $H^3(X,\IZ)$ into $\alpha_A$ and $\beta^A$, where
\eqn\awb{\int\alpha_A\wedge\beta^B = \delta_A{}^B.}

For the Calabi-Yau manifold $X$, all of these basis forms are closed,
by the definition of cohomology.  To geometrically twist $X$, we
introduce relations analogous to the defining closure relation for a
twisted torus: $d\eta^m + \ha\gamma^m{}_{np}\eta^n\wedge\eta^p=0$.
Since we have no global 1-forms on a Calabi-Yau, the geometric twist
must be expressed in terms of $\omega_a$, $\tilde\omega^a$, $\alpha_A$
and $\beta^A$.

Suppose that we modify the Calabi-Yau cohomology by introducing the
twisted closure relations
\eqn\dtwo{d\omega_a = -M_a{}^A\alpha_A + N_{aA}\beta^A,}
while retaining the condition that $\alpha_A\wedge\omega_a$ and
$\beta^A\wedge\omega_a$ be cohomologically trivial.  Here, $M_a{}^A$
and $N_{aA}$ are $b_2\times\half b_3$ integer matrices (up to an
overall scale factor, cf.~Sec.~3.3), where $b_n$ denotes the $n$th
Betti number.  Then,
$$\eqalign{N_{aA} &= \int\alpha_A\wedge d\omega_a
= -\int (d\alpha_A)\wedge\omega_a,\cr
M_a{}^A &= \int\beta^A\wedge d\omega_a
= -\int (d\beta^A)\wedge\omega_a.}$$
From the second equality in each line,
\eqn\dthree{d\alpha_A = -N_{aA}\tilde\omega^a\quad{\rm and}\quad
d\beta^A = -M_a{}^A\tilde\omega^a,}
respectively.  So, the twisting of $\alpha_A$ and $\beta^A$ is
determined from that of $\omega_a$ without any additional data.  The
consistency condition $d^2\omega_a = 0$ is
\eqn\ddvanish{NM^T - MN^T=0.}
As already mentioned in the introduction, while this prescription for
discrete deformation of the Calabi-Yau cohomology is straightforward,
it does not amount to a construction of a new non Calabi-Yau manifold
$\tilde X$ realizing the modified cohomology groups.  It is not clear
that such a manifold actually exists for arbitrary choice of discrete
twist data $M_a{}^A$ and $N_{aA}$ satisfying Eq.~\ddvanish.  A goal of
this paper is to explicitly construct $\tilde X$ for the Voisin-Borcea
class of Calabi-Yau manifolds $X$ (with special emphasis on $\TZZ$)
and for a subset of twists $M_a{}^A$ and $N_{aA}$.
 
Note that we can express Eqs.~\dtwo\ through \ddvanish\ in a
manifestly symplectically covariant manner if we define a vector
$$\alpha = \pmatrix{\alpha_A\cr\beta^A},$$
and matrices
$${\bf M} = \pmatrix{N_{aA} &M_a{}^A},\qquad
{\bf\Lambda} = \pmatrix{0 & -\delta_A{}^B\cr \delta_A{}^B & 0}.$$
Then Eqs.~\dtwo\ and \dthree\ can be written
$$d\omega = {\bf M \Lambda}\alpha\quad{\rm and}\quad
d\alpha^T = \tilde\omega^T{\bf M}.$$
The consistency condition $d^2\omega_a = 0$ of Eq.~\ddvanish\ becomes
${\bf M}{\bf\Lambda}{\bf M}^T = 0$.

It is clear that the geometric twists lift a part of the cohomology
ring of the original Calabi-Yau manifold $X$, since the forms on the
RHS of Eqs.~\dtwo\ and \dthree\ are now exact, while those appearing
on the LHS now fail to be closed.  The precise statement is that the
free (non-torsion) part of the cohomology is reduced from
\def\rankM{{{\rm rank}{(\bf M)}}}
$$H_\free^3(X) = \IZ^{b^3(X)}\quad{\rm and}\quad H_\free^4(X) =
\IZ^{b^2(X)}\quad\hbox{for the Calabi-Yau},$$
to
$$H_\free^3(\tilde X) = \IZ^{b^3(X)-\rankM}
\quad{\rm and}\quad 
H_\free^4(\tilde X) = \IZ^{b^2(X)-\rankM}$$
after introducing the geometric twists.  Assuming that the original
Calabi-Yau manifold $X$ has trivial torsion, the torsion part of the
cohomology of the twisted manifold $\tilde X$ is\foot{This result uses
Poincar\'e duality and the Universal Coefficient Theorem \BottTu\
$H^{n+1}_\tor = H^\tor_n$.}  $H^3_\tor(\tilde X) = H^4_\tor(\tilde X)
=
\bigotimes_{i = 1}^\rankM \IZ_{m_i}$, where the $m_i$ are determined
by the particular choice of the matrix ${\bf M}$.

The interpretation of the Calabi-Yau cohomology $H^*(X,\IZ)$ from the
point of view of the twisted manifold $\tilde X$ is as a ``first
approximation'' to the cohomology $H^*(\tilde X,\IZ)$ in a sense that
has been made precise by Tomasiello~\TomasielloBP\ using spectral
sequences.  In this description, the cohomology of $X$ corresponds to
the term $E^{p,q}_2$ in a Leray-Hirsch spectral sequence.  After a
finite number of steps (which in this context we expect to be just one
more) the sequence converges to the fixed value $E^{p,q}_\infty$,
which gives the cohomology of $\tilde X$.


\newsec{Geometric twists of $\TZZ$}

We now specialize to the Calabi-Yau orbifold $X=\TZZ$.  There are two
versions of this orbifold and we restrict to the choice with trivial
discrete torsion.  This choice gives topology $(h^{1,1},h^{2,1}) =
(51,3)$.  For the other choice, the Hodge numbers are reversed.  

The two $\IZ_2$ involutions each invert a $T^4$ within the $T^6$:
\eqn\Ztwos{\eqalign{%
\s_3\colon\quad& (x^1,x^2;x^3,x^4;x^5,x^6)\to
(-x^1,-x^2;-x^3,-x^4;+x^5,+x^6),\cr
\s_1\colon\quad& (x^1,x^2;x^3,x^4;x^5,x^6)\to
(+x^1,+x^2;-x^3,-x^4;-x^5,-x^6).}}
The composition $\s_2 = \s_3\otimes\s_1$ inverts a third $T^4$.  Let
$T^2_{(\a)}$ denote the torus covered by coordinates
$(x^{2\a-1},x^{2\a})$ for $\a=1,2,3$; and let $T^4_{(\a)}$ be the
corresponding 4-torus covered by the complementary set of coordinates.
The involutions have been labeled so that $\s_\a$ leaves $T^2_{(\a)}$
invariant, but inverts $T^4_{(\a)}$.

\subsec{Topology of $\TZZ$}

A part of the (co)homology of $X=\TZZ$ is inherited directly from the
$T^6$.  Let us focus on the divisors.  The subgroup of $H_4(X,\IZ)$
inherited from the $T^6$ is generated by ``sliding divisors'' made up
of $\IZ_2$ invariant pairs of 4-tori on the $T^6$:
\eqn\Falpha{2F_\a = T^4_{(\a)}\times\{(a,b)\cup (-a,-b)\}
\quad\hbox{for $(a,b)\in T^2_{(\a)}$ not a $\IZ_2$ fixed point.}}
(The factor of 2 anticipates that $F_\a$ is also an element of
$H_4(X,\IZ)$ as we discuss below.)  The Poincar\'e dual cohomology
generators are $2\o_\a$, where
\eqn\Xsecondcohom{\omega_1 = 2dx^1\w dx^2,\quad 
\omega_2 = 2dx^3\w dx^4, \quad\omega_3 = 2dx^5\w dx^6.}
In addition, there are 48 exceptional divisors $E_{\a I}$ with dual
cohomology generators
\eqn\Xsecondcohomtw{\o_{\a I},\quad\hbox{where}\quad\a=1,2,3,\quad I =
1,\ldots,16.}
These come from blowing up the $3\times16$ fixed lines $\IP^1 =
T^2_{(\a)}/\IZ_2$ located at transverse coordinates each equal to 0 or
1/2.

The classes $2F_\a$ and $E_{\a I}$ with integer coefficients, generate
only a subgroup of $H_2(X,\IZ)$.  This subgroup omits some of the
divisors that arise in orbifold twisted sectors.\foot{Unfortunately,
the word ``twisted'' can refer to either a winding sector in an
orbifold conformal field theory, or to a topology that has been
discretely modified, i.e., in going from a product space to a
fibration.  Whenever we refer the former, we will use the words
orbifold twisted sector.} We now describe these divisors as well.

First let us consider those divisors that arise in the orbifold
twisted sector with respect to only a $\IZ_2$ subgroup of
$\IZ_2\times\IZ_2 = \{1,\s_1,\s_2,\s_3\}.$ Performing the orbifold
quotient in two steps gives three presentations of $X$ of the form
$(K3\times T^2)/\IZ_2$, one for each of the three choices $K3_{(\a)} =
T^4_{(\a)}/\sigma_\a$.  The $\a$th presentation makes it clear that
$K3_{(\a)}$ exists as a cycle in the Calabi-Yau manifold $X$ as the
generic fiber over base $\IP^1 = T^2_{(\a)}/\IZ_2$.  The homology
class of this divisor is half of the class of $T^4_{(\a)}$, i.e.,
$F_\a$.

The classes $F_\a$ and $E_{\a I}$ with integer coefficients still
generate only a subgroup of $H_2(X,\IZ)$.  The remaining divisors are
as follows. Before resolving the fixed points, $\TZZ$ has $3\times4$
divisors of topology $\IP^1\times\IP^1$ from the fixed planes
$T^4_{(\a)}/(\IZ_2\times\IZ_2)$ located at each of the 4 fixed points
in the transverse coordinates.  These divisors persist after the
resolution and represent homology classes of the form
$$\half F_\a - \half(\hbox{sum of eight }E_{\b I}),$$
as is discussed in more detail in in App.~B.  These divisors together
with the $F_\a$ and $E_{\a I}$ generate all of $H_2(X,\IZ)$.

For $H^3(X,\IZ)$, the story is simpler.  The only complex structure
deformations of $\TZZ$ are those inherited from the three
$T^2_{(\a)}$.  The forms
\eqn\Xthirdcohom{2dx^i\w dx^j\w dx^k,\quad\hbox{where}\quad 
i=1,2,\quad j=3,4,\quad k=5,6.}
on $T^6$ give a basis for $H^3(X,\IZ)$.  However, it should be noted
that the ``sliding 3-cycles'' inherited from $T^6$ consist of
$\IZ_2\times\IZ_2$ invariant quadruples of 3-tori in $T^6$.  The dual
cohomology classes are twice those that appear in Eq.~\Xthirdcohom.

The presence of the $3\times 16$ exceptional cycles $E_{\a I}$ and
absence of exceptional complex structure deformations can be
understood intuitively as follows.  As discussed above, there are
three ways to view the orbifold $X$ as $(K3\times T^2)/\IZ_2$,
corresponding to the three choices $K3_{(\a)} = T^4_{(\a)}/\sigma_\a$.
Each $K3_{(\a)}$ has 16 exceptional cycles, corresponding to 16
hyperK\"ahler deformations that smooth the singularities of
$T^4_{(\a)}/\IZ_2$.  The $\omega_{\a I}$, for $I = 1,\ldots,16,$
generate these hyperK\"ahler deformations of $K3_{(\a)}$.  Finally,
each hyperK\"ahler deformation of $K3$ is generated by three real
moduli.  Given a choice of complex structure on the $K3$, two of these
moduli generate complex structure deformations and the remaining
modulus generates a K\"ahler deformation. When we perform the second
$\IZ_2$ operation, the explicit $\IZ_2$ of $(K3\times T^2)/\IZ_2$, we
project out the exceptional complex structure deformations of the
$K3_{(\a)}$ and retain the K\"ahler deformations.  This leads to Hodge
numbers $h^{1,1} = 3+3\times16=51$ and $h^{2,1} = 3$, in agreement
with Eqs.~\Xsecondcohom, \Xsecondcohomtw\ and \Xthirdcohom.


\subsec{Geometric twists in the orbifold untwisted sector}

The only geometric twists of $X=\TZZ$ that have been discussed in the
literature to date are those that are inherited from the twisted
$T^6$.  The twisted $T^6$ is a parallelizable six-manifold with global
1-forms $\eta^m$, $m=1,\ldots,6$ (sections of the frame bundle),
satisfying $d\eta^m+\ha\g^m_{np}\eta^n\w\eta^p=0$.  As the name
suggests, this can be viewed as a discretely deformed version of the
ordinary $T^6$, characterized by the twist data $\g^m_{np}$
antisymmetric in lower indices.\foot{For the ordinary $T^6$, we have
$\g^m_{np} = 0$ and sections of the frame bundle are constant linear
combinations of the closed coordinate 1-forms $dx^m$.} For the
geometry to be well-defined, the $\g^m_{np}$ must satisfy $\g^m_{mn}=
0$ (for the existence of a global volume form) together with the
Jacobi identity (for $d^2=0$), as consequence of which they define the
structure constants of a Lie algebra.  If the corresponding Lie group
is compact, then the twisted $T^6$ is defined by this Lie group.  If
it is noncompact, then subject to certain existence caveats, the
twisted $T^6$ is defined as the coset of the Lie group by a discrete
subgroup.

We now relate the $\g^m_{np}$ to the matrices $M$ and $N$ of the
previous section.  As above, define $T^2_{(\a)}$ to be the torus
covered by coordinates $(x^{2\a-1},x^{2\a})$ for $\a=1,2,3$.  The
components of $\g^m_{np}$ that survive the orbifold projection are
those with $m,n,p$ each on a different $T^2_{(\a)}$, and those with
all three indices on the same $T^2_{(\a)}$.  For simplicity, we set
the latter components to zero; then the constraint $\g^m_{mn}=0 $ is
trivially satisfied.  Finally, we define twisted analogs of the
differential forms \Xsecondcohom\ and \Xthirdcohom.  These are
globally defined differential forms on $\tilde X 
= (\hbox{twisted-}T^6)/\IZ_2\times\IZ_2$:
\eqn\twofourforms{\eqalign{%
\o_1 &= 2\eta^1\w \eta^2,\cr
\o_2 &= 2\eta^3\w \eta^4,\cr
\o_3 &= 2\eta^5\w \eta^6,}
\quad\eqalign{%
\tilde\o^1 &= 2\eta^3\w\eta^4\w\eta^5\w\eta^6,\cr
\tilde\o^2 &= 2\eta^1\w\eta^2\w\eta^5\w\eta^6,\cr
\tilde\o^3 &= 2\eta^1\w\eta^2\w\eta^3\w\eta^4,}}
and
\eqn\threeformsab{\eqalign{%
\a_0 &= 2\eta^1\w\eta^3\w\eta^5,\cr
\a_1 &= -2\eta^1\w\eta^4\w\eta^6,\cr
\a_2 &= -2\eta^2\w\eta^3\w\eta^6,\cr
\a_3 &= -2\eta^2\w\eta^4\w\eta^5,}
\quad\eqalign{%
\b^0 &= -2\eta^2\w\eta^4\w\eta^6,\cr
\b^1 &= 2\eta^2\w\eta^3\w\eta^5,\cr
\b^2 &= 2\eta^1\w\eta^4\w\eta^5,\cr
\b^3 &= 2\eta^1\w\eta^3\w\eta^6.}}

In terms of $\g^i_{jk}$, the matrices $M$ and $N$ are
\eqn\Mgamma{M_a{}^A = \pmatrix{%
\g^2_{53} & -\g^2_{64} & \g^1_{63} & \g^1_{54}\cr
-\g^4_{51} & -\g^3_{61} & \g^4_{62} & -\g^3_{52}\cr
\g^6_{31} & \g^5_{41} & \g^5_{32} & -\g^6_{42}},
\quad
N_{a A} =\pmatrix{%
-\g^1_{64} & \g^1_{53} & -\g^2_{54} & -\g^2_{63}\cr
\g^3_{62} & \g^4_{52} & -\g^3_{51} & \g^4_{61}\cr
-\g^5_{42} & -\g^6_{32} & -\g^6_{41} & \g^5_{31}}.}

The consistency condition~\ddvanish\ can be satisfied by choosing, for
example, $M_\a{}^A=0$.  In the case of intersecting D-brane models,
this is a requirement rather than a choice, as $\o_a$, $\tilde\o^a$
and $\b^A$ are odd while $\a_A$ is even under the orientifold $\IZ_2$
operation $\O (-1)^{F_L}\CI_3$.  Here, $\O$ is worldsheet parity,
$(-1)^{F_L}$ is left-moving fermion parity and
\eqn\OrientInv{\CI_3\colon\quad(x^1,x^2;x^3,x^4;x^5,x^6)
\to (x^1,-x^2;x^3,-x^4;x^5,-x^6).}


\subsec{Geometric twists in the orbifold twisted sector}

In this section, we show how to introduce geometric twists involving
16 of the 48 blow-up modes of $X=\TZZ$ through the following trick:
The 16 exceptional K\"ahler deformations of $K3 = T^4/\IZ_2$ are in
the untwisted sector of $(K3\times T^2)/\IZ_2$, but in the twisted
sector of $\TZZ$.  Therefore, if we twist $K3\times T^2$ by fibering
the $K3$ surface over the $T^2$ while at the same time preserving the
existence of a $\IZ_2$ involution, then the $\IZ_2$ quotient of the
result is a twisted version of $\TZZ$.  For generic fibration, the
twists involve all 16 of the $\TZZ$ blow-up modes inherited from the
$K3.$

\nonnumberedsubsec{Moduli space of K3}

To describe a smooth fibration of $K3$ over $T^2$ we simply allow
the moduli of the $K3$ surface to vary over $T^2$.  The moduli
space of metrics on $K3$ is the coset
\eqn\ModKthree{\CM = \IR_{>0}\ \times\ {} 
\bigl(SO(3)\times SO(19)\bigr)\backslash SO(3,19)/\G_{3,19}.}
Here, the first factor in Eq.~\ModKthree\ is the overall volume of the
$K3$ surface.  The second factor is the choice of hyperK\"ahler
structure. Its explicit form arises as follows. The coset describes
the space of positive signature 3-planes in $\IR^{3,19}$.  This space
appears since the cohomology group $H^2(K3,\IR)$ has signature
$(3,19)$ with respect to the inner product $(\o_1,\o_2) =
\int\o_1\w\o_2$.  That is, there are three selfdual 2-forms and
nineteen anti-selfdual \hbox{2-forms} on $K3$.  The choice of
hyperK\"ahler structure on $K3$ is the choice of positive 3-plane
spanned by $J$, $\Re\O_{(2)}$, $\Im\O_{(2)}$ in $H^{2}(K3,\IR)$.
Finally, the quotient by the discrete group $\G_{3,19}$ accounts for
the fact that automorphisms of the lattice $H^2(K3,\IZ)$ relate
equivalent $K3$ surfaces.

The $\bigl(SO(3)\times SO(19)\bigr)\backslash SO(3,19)/\G_{3,19}$ coset can be
parametrized as \MaharanaMY
\eqn\CosetRep{M^{ab} = \pmatrix{%
    G & -G C                 & -G A^T\cr
-C^TG & G^{-1} + a + C^T G C & A^T + C^T G A^T\cr
-A G  & A + A G C            & 1 + A G A^T},}
where $a^{ij}= A^{Ii} A^{Ij}$ and $C^{ij} = B^{ij}+\ha A^{Ii} A^{Ij}$.
Here, $G_{ij}$ is a metric on $T^3(x^1,x^2,x^3)$, with coordinate
periodicity $x^i\simeq x^i+1$; $B^{ij}$ is an antisymmetric bivector
$T^3$ with periodicity $B^{ij}\simeq B^{ij}+1$; and $\half A^{Ii}$,
$I=1\ldots16$, are the coordinates of sixteen points on the $T^3$.
The index upper $a$ runs over lower $i$, upper $i$ and $I$.  In
addition to the periodicities listed, and of course the $SL(3,\IZ)$
equivalence of $G_{ij}$ under change of $T^3$ lattice basis, there are
further identifications of $G,B,A$ under more general elements of
$\G_{3,19}$.

This coset description is well known and roughly speaking makes
manifest the duality between M theory on $K3$ and the heterotic or
type I string on $T^3$.  To be more precise, given our definitions of
$G,B,A$ (note the index placement, in particular), the natural duality
relates M theory on $K3$ not to type I, but rather to type
I${}'''$---the T-dual of type I on $T^3$ under inversion of all three
directions of the $T^3$.  In this version of the duality, the $K3$
moduli $\half A^{Ii}$ are identified with the transverse scalars to
the sixteen D6-branes, and the $K3$ moduli $B^{ij}$ are identified
with $-\half\e^{ijk}C_{(1)k}$, where $C_{(1)}$ is the the RR 1-form.
We have chosen to parametrize the $K3$ moduli space in terms of
scalars adapted to this duality frame rather than the heterotic/type I
duality frame for the following reason: The torus twists $\g^i_{jk}$
discussed in Sec.~3.1 correspond to twists of the physical $T^3$ in
type I${}'''$ variables, but to twists of the dual $T^3$ in the
heterotic/type I variables
\SchulzKK.

Among the possible presentations of $SO(3,19)$, the coset
representative $M^{ab}$ leaves invariant the $SO(3,3+16)$ metric in
off-diagonal form,
\eqn\SOmetric{\eta_{ab} = \pmatrix{%
0 & \d^i{}_j & 0\cr
\d_i{}^j & 0 & 0\cr
0 & 0 & \d_{IJ}}.}
\noindent It can be expressed concisely as $M = V^TV$, where $V_\L{}^a$ is
the vielbein
\eqn\SOvielbein{V(E,B,A)=\pmatrix{E & -E C &-E A^T\cr
0 & E^{-1T} & 0\cr 0 & A & 1},}
and $E^\l{}_i$ is the vielbein for the metric $G_{ij}$ on $T^3$.
Here, $\L$ is the $SO(3)\times SO(19)$ coset index and $a$ is the
$SO(3,19)$ index.  If we choose to write $A,B,C$ with vielbein indices
instead of coordinate indices, then the last expression instead takes
the form
\eqn\Altvielbein{\tilde V(E,B,A) =\pmatrix{E & -CE^{-1T} & -A^T\cr
0 & E^{-1T} & 0\cr 0 & AE^{-1T} & 1}.}

\nonnumberedsubsec{Cohomology of $K3$}

The second cohomology group of $K3 = T^4/\IZ_2$ is generated by the
subgroup inherited from $T^4$,
\eqn\Kthreecohom{\chi^i =  2dx^4\w dx^i\quad\hbox{and}\quad
\chi_i = \e_{ijk}dx^j\w dx^k,\quad i,j,k=1,2,3,}
together with the blow-up modes
\eqn\Kthreeblowup{\chi_I,\quad I = 1,\ldots,16,}
dual to the exceptional curves that arise from blowing up the sixteen
$\IZ_2$ fixed points.\foot{In our basis, $dx^4$ plays a privileged
role since we work in the convention that $e^4$ is the real part of
the hyperK\"ahler 1-form on $T^4$.  The forms $e^{1,2,3}$ are the
three (hyper)imaginary parts.  See App.~A for definitions of the
$e^i$ and a review of hyperK\"ahler structure on $T^4$.}  The
cohomology classes $\chi_a = (\chi^i,\chi_i,\chi_I)$ satisfy
\eqn\chimatrix{\int_{K3}\chi_a\w\chi_b = -2\eta_{ab},}
where $\eta_{ab}$ is the $SO(3,3+16)$ invariant metric introduced in
Eq.~\SOmetric.\foot{The change of basis from $\chi_a$ with
intersection matrix $-2\eta_{ab}$ to $\chi'_a$ with the intersection
matrix $-\hbox{Cartan}(E_8\times E_8)\times U_{1,1}^3$ (referred to
later in the Voisin-Borcea context) can be found in App.~B of
Ref.~\HeadrickCH.  Unlike the $\chi_a$, which require some
half-integer coefficients to generate all of $H^2(K3,\IZ)$, the
$\chi'_a$ generate $H^2(K3,\IZ)$ with purely integer coefficients.  We
thank K.~Wendland for providing this reference.}

Much like in the $\TZZ$ discussion in Sec.~3.1, the classes
\Kthreecohom\ and \Kthreeblowup, with integer coefficients, do not
generate all of $H^2(K3,\IZ)$, but rather an order two sublattice
called the Kummer lattice.  To obtain the full $H^2(K3,\IZ)$ we also
need to include generators of the form
$$\half f - \half(\hbox{sum of four }\chi_I),
\quad\hbox{where $f=\chi^i$ or $\chi_i$,}$$
dual to certain $\IP^1$ divisors of $K3$.  This is discussed in more
detail in App.~B.

\nonnumberedsubsec{Cohomological interpretation of the coset matrix}

The inverse $M_{ab}$ of the coset representative $\CosetRep$ has a
simple interpretation in terms of the natural metric-dependent inner
product on the $\chi_a$:
\eqn\chichiM{\int_{K3}\star\chi_a\w\chi_b = 2M_{ab}.}
If we define $\chi_\L = V_\L{}^a\chi_a$, then from the definition of the
vielbein $V_\L{}^a$, we have
\eqn\chiLchiL{\int_{K3}\star\chi_\L\w\chi_{\L'} = 2\d_{\L\L'}.}
So, the vielbein $V_\L{}^a$ takes the moduli-independent cohomology
basis $\chi_a$ with moduli-dependent norm into a moduli-dependent
basis $\chi_\L$ with moduli-independent norm.\foot{The attentive
reader may have noticed that setting $A^{Ii}_n=0$ in the coset matrix
\SOvielbein\ should correspond to the $SO(32)$ point in moduli space.
However, given our definitions, we obtain the $(A_1)^{16}$ point of
the unresolved $T^4/\IZ_2$ instead.  The latter should corresponds to
evenly distributing the 16 $A^{Ii}$ among the the 8 fixed values where
each of the $i=1,2,3$ components is either $0$ or $1/2$.  This problem
is solved by first performing a change of basis $\chi\to \hat\chi =
V(1,1,-\D A)\chi$, in order to convert the $(A_1)^{16}$ adapted basis
to the $SO(32)$ adapted basis, and then writing $\chi_\L =
V_\L{}^a\hat\chi_a$.  For notational simplicity, we leave this change
of basis implicit.}

\nonnumberedsubsec{Periodicities of $K3$ moduli}

Define the $SO(3,19)$ elements
\eqn\Vonetwothree{V_1(x) = \pmatrix{x & 0 & 0\cr
0 & x^{-1T} & 0\cr 0 & 0 & 1},
\ V_2(y)=\pmatrix{1 & -y & 0\cr
0 & 1 & 0\cr 0 & 0 & 1},
\ V_3(z)=\pmatrix{1 & -\half z^T z & -z^T\cr
0 & 1 & 0\cr 0 & z & 1}.}
Then, the vielbein $V(E,B,A)$ and $\tilde V(E,B,A)$ defined above are
\eqn\Vdecomp{V = V_1(E)V_2(B)V_3(A),\quad\tilde V = V_3(A)V_2(B)V_1(E).}
From the interpretation $\chi_\L = V_\L{}^a\chi_a$ given in the last
subsection, we obtain identifications of the $K3$ moduli from the
automorphisms $\G_{3,19}$ of $H^2(K3,\IZ)$.  The group $\G_{3,19}$ is
a discrete subgroup of $SO(3,19)$, which in turn contains, as a proper
subgroup, the group $SO(3,19,\IZ)$ of integer matrices $\G_a{}^b$
preserving the intersection matrix \chimatrix.  (Since the $\chi_a$
require some half integer coefficients to generate all of
$H^2(K3,\IZ)$, there are also elements of $\G_{3,19}$ which contain
half integer components, and are therefore not contained in
$SO(3,19;\IZ)$.)  Let us focus on this subgroup.  For any such
automorphism $\G_a{}^b\in SO(3,19;\IZ)$ we obtain an identification
$V(E,B,A)_\L{}^b\simeq V(E,B,A)_\L{}^a\G_a{}^b$.  From $V\simeq VV_3$
with $V_3(z)\in SO(3,19;\IZ)$, we obtain the identifications
$A^{iI}\simeq A^{iI}+2$ for individual components $A^{Ii}$.  Beyond
this, for $n$-tuples of components there exist additional
identifications, for example $(A^{Ii},A^{Ji})\simeq
(A^{Ii}+1,A^{Ji}+1)$.  From $V\simeq VV_2$ with $V_2(y)\in
SO(3,19;\IZ)$ we obtain the identifications $B^{ij}\simeq B^{ij}+1$.

\nonnumberedsubsec{Fibration of $K3$ over $T^2$}

In Sec.~3.2, we defined a twisted torus to be a parallelizable
manifold, analogous to a torus, but with global 1-forms $\eta^m$
satisfying $d\eta^m + \half\g^m_{np}\eta^n\w\eta^p=0$, which can be
viewed as generalizations of the coordinate 1-forms $dx^m$ on a torus.
A special case of a twisted torus is a torus fibration over torus
base.  In this case, let us refine the index notation that we have
been using, so that $i,j,\ldots$ denote fiber indices and $m,n,\ldots$
denote base indices.  Then, for the fibration, we take $\eta^n = dx^n$
on the base, and have
\eqn\twToverT{d\eta^i + \g^i_{nj} dx^n\w\eta^j=0,}
for the 1-forms on the fiber.

Analogously, for a $K3$ fibration over $T^2(x^5,x^6)$, the global
2-forms on the fibration include twisted versions of the $\chi_a$,
which we will denote by $\z_a$.  The $\z_a$ satisfy
\eqn\dzeta{d\zeta_a + M_{na}^b dx^n\w\zeta_b = 0.}
From the topological consistency condition $d^2=0$, we require
$[M_m,M_n] = 0$.  Let us define $\G_5 = \exp(-x^5 M_5)$ and $\G_6 =
\exp(-x^6 M_6)$.  The closure condition $\dzeta$ follows by
promoting the forms $\chi_a$ on $K3$ to global forms on the
fibration through the relation
\eqn\zetachi{\zeta_a = \bigl(\G_5(x^5)\G_6(x^6)\bigr)_a{}^b\chi_b.}
In order to preserve the inner product $\chimatrix$ and integrality of
the basis, the monodromy matrices $\G_n(1)$ should be elements of
$\G_{3,19}$.  The monodromy matrices, or equivalently, the Lie algebra
elements $M_5$ and $M_6$, completely determine the topology of the
fibration.

Let us write
\eqn\GammaEBA{\G_5(x^5) = \tilde V(\eta_{(5)},\b_5 x^5, m_5 x^5)
\quad\hbox{and}\quad
\G_6(x^6) = \tilde V(\eta_{(6)},\b_6 x^6, m_6 x^6),}
where vielbein $\tilde V$ was defined in Eq.~\Altvielbein.  This
ansatz describes the subset of $K3$ fibrations for which the $K3$
moduli $B^{ij},A^{Ii}$ undergo periodic linear shifts as we traverse
the $x^5$ and $x^6$ circles on the $T^2$ base. (Here, the reader may
wish to refer to the discussion above on the cohomological
interpretation of the coset vielbein).  Here, $\beta^{ij}_n$ and
$m^{Ii}_n$ are constant matrices, and the $\eta^i_{(n)j}(x)$ are given
in terms of constants $\g^i_{5j}$ and $\g^i_{6j}$ as
\eqn\EtaFromGamma{\eta^i_{(5)j} = \exp(-\gamma^i_{5j} x^5)
\quad\hbox{and}\quad
\eta^i_{(6)j} = \exp(-\gamma^i_{6j} x^6).}
Let us also write $\eta^i = \eta^i_{(5)j}\eta^j_{(6)k}dx^k$.  Then,
the $\eta^i$ satisfy Eq.~\twToverT, provided that $\g_5$ and $\g_6$
commute.  And indeed, this is the case: $[\g_5,\g_6]=0$, as a
consequence of $[M_5,M_6]=0$ (cf.~the first condition in Eq.~(3.27)
below).

In our discussion of the moduli space of $K3$, the coset matrix
$M^{ab}$ was defined in terms of tensors on a formal auxiliary $T^3$.
We can think of the $\eta^i$ that we have just defined as a frame for
a fibration of this formal $T^3$ over the physical $T^2$ that forms
the base of the $K3$ fibration.  The $\eta^i$ encode the dependence of
the $K3$ moduli $G_{ij}$ on the base coordinates $x^5,x^6$, as
quantified by the data $\g^i_{nj}$.  Likewise, $\b^{ij}_n$ and
$m^{Ii}_n$ parametrize the base coordinate dependence of the remaining
$K3$ moduli $B^{ij}$ and $A^{Ii}$, respectively.

Eqs.~\zetachi\ and \GammaEBA\ together give the change of basis from
local untwisted forms $\chi_b$ to the global twisted forms $\zeta_a$
on the $K3$ fibration.  The reason that we have used the presentation
$\tilde V$ rather than $V$ in \GammaEBA\ is that it simplifies the
results below if we define $\b^{ij}_n$ and $m^{iI}_n$ to be tensor
components with respect to the frame $\eta^i$ rather than the
coordinate 1-forms $dx^i$.

By differentiating Eq.~\GammaEBA, we obtain
\eqn\Mexpression{M_n =
\pmatrix{\g_n & \b_n & m^T_n\cr
0 & -\g^T_n & 0\cr
0 & -m_n & 0},}
which determines the closure condition $\dzeta$.  In components,
Eq.~\dzeta\ reads
\eqn\dzexplicit{\eqalign{%
d\z^i &= -dx^n\w(\g^i_{nj}\z^j +\b^{ij}_n\z_j + m^{Ii}_n\z_I),\cr
d\z_i & = dx^n\w\g^j_{ni}\z_j,\cr
d\z_I & = dx^n\w \d_{IJ}m^{Ji}_n\z_i,}}
and the condition $[M_n,M_p]=0$ becomes
\def\vpT{{\vphantom{T}}}
\eqn\geobianchi{\g^\vpT_{[n}\g^\vpT_{p]} = 0,\quad
-\g^\vpT_{[n}\b^\vpT_{p]}+\b^\vpT_{[n}\g^T_{p]} + m^T_{[n}m^\vpT_{p]} = 0,
\quad \g^\vpT_{[n}m^T_{p]} = 0.}

\nonnumberedsubsec{Quantization and elliptic versus parabolic twists}

To define a $K3$ fibration over $T^2$, the twist data $\g^i_{nj}$,
$\b^{ij}_n$ and $m^{Ii}_n$ must be chosen so that the monodromy
matrices $\G_5 = \exp(-x^5M_5)$ and $\G_6 = \exp(-x^6M_6)$ are
elements of $\G_{3,19}$.  In deriving the result~\Mexpression, we have
further restricted to a particularly simple subgroup of $SO(3,19;\IZ)$
corresponding to the subgroup of $SO(3,19)$ spanned by our choice of
vielbein for the $\bigl(SO(3)\times SO(19)\bigr)\backslash SO(3,19)$
coset representatives.  For this subgroup, we now relate the condition
$\G_5,\G_6\in SO(3,19;\IZ)$ to restrictions on the allowed twist data.

For the $\b^{ij}_n$ and $m^{Ii}_n$, there is no subtlety.  The
conditions $\b^{ij}_n\in\IZ$ and $m^{Ii}_n\in 2\IZ$ lead to integer
components in $\G_5,\G_6$, and follow directly from the periodicities
$B^{ij}\simeq B^{ij}+1$ and $A^{iI}\simeq A^{iI}+2$.  For the
$\g^i_{nj}$, consider the diagonal blocks $\eta_{(n)}$ and
$\eta_{(n)}{}^{-1T}$ of $\G_n$ (cf.~Eq.~\EtaFromGamma).  The
constraint that these be integer matrices gives rise to qualitatively
different conditions, depending on the nilpotency properties of the
$\g_n$ (or equivalently, the idempotency properties $\eta_{(n)}$).  We
will consider only two special cases.

First, consider the case that $(\g_{(n)})^2=0$.  We will refer to this
as the ``parabolic case'' by analogy to the conjugacy classes of
$SL(2,\IR)$.\foot{The hyperbolic, elliptic and parabolic conjugacy
classes of $SL(2,\IR)$ can be represented by matrices
$\G=\bigl({\exp(-x)\atop 0}{0\atop\exp(x)}\bigr)$,
$\bigl({\cos(x)\atop-\sin(x)} {\sin(x)\atop\cos(x)}\bigr)$ and
$\bigl({1\atop0}{x\atop1}\bigr)$, which are the exponentials $\exp(-\g
x)$ for $\g=\bigl({1\atop 0}{0\atop-1}\bigr)$,
$\bigl({0\atop1}{-1\atop0}\bigr)$ and
$\bigl({0\atop0}{1\atop0}\bigr)$, respectively.  See
Ref.~\DabholkarSY\ for a discussion of Scherk-Schwarz
compactifications with $SL(2,\IZ)$ monodromy over $S^1$.}  Then
$\eta^i_{(5)j} = 1-\g^i_{5j}x^5$ and $\eta^i_{(6)j} = 1-\g^i_{6j}x^6$.
In order that these be integer matrices at $x^n=1$, we require
$\g^i_{nj}\in\IZ$.

In contrast, consider the case that $\g_{(n)}$ is not nilpotent of any
degree, but instead $\eta_{(n)}$ satisfies $(\eta_{(n)})^k =1$ for
some finite positive integer $k$.  We will refer to this as the
``elliptic case,'' again by analogy to $SL(2,\IR)$.  For example,
suppose that $\g_{(n)} = (\pi\l_{(n)}/2)\CI$, for some integer matrix
$\CI$ satisfying $\CI^2 = -1$.  Here $\l_{(n)}$ is a proportionality
constant.  Then,
\eqn\EllipticEta{\eta_{(5)}(x^5) = \exp(-\g_5 x^5) =
\cos(\smallfrac\pi2\l_{(5)}x^5) -\CI\sin(\smallfrac\pi2\l_{(5)}x^5),}
with a similar expression for $\eta_{(6)}$.  In this case, it is clear
that for $\eta_{(n)}$ to be integer valued at $x^n = 1$, we require
$\l_{(n)}\in\IZ$, and hence $\gamma^i_{nj}\in\smallfrac\pi2\IZ$.  Note
that $\eta_{(n)}$ is then idempotent of degree four,
$(\eta_{(n)})^4=1$.

The qualitative difference between the two cases is that the first
requires energy, i.e., curvature, since some of the K3 moduli vary
linearly as a function of the base coordinates.  This case therefore
gives rise to a fibration of $K3$ over $T^2$ that metrically is
locally distinguishable from a direct product.  The second case
typically requires some of the $K3$ moduli to be fixed to $\IZ_k$
symmetric values.  This does not require energy from spatially varying
moduli, or curvature.  So, in this case, there is a global
distinction, but no local metric distinction between the fibration and
$K3\times T^2$.  Unless we state otherwise, we will have the former,
parabolic case in mind in the remainder of the paper.

\nonnumberedsubsec{Fibration of $K3$ over $T^2$, with $\IZ_2$
involution}

Consider the $\IZ_2$ involution $\s_1$ acting on $K3_{(3)}\times
T^2_{(3)}$, and let us focus on this $K3$.  The differential forms
$\chi_3$ and $\chi^3$ are even under the action of $\s_1$, while
$\chi_{1,2}$ and $\chi^{1,2}$ are odd.  Since we defined $\s_1$ in the
context of $\TZZ$, the K3 blow-up modes $\chi_I$ are all even under
$\s_1$ and all survive as K\"ahler moduli in the quotient Calabi-Yau.
(For Voisin-Borcea Calabi-Yau manifolds based on other $K3$
involutions, the $\chi_I$ in general split into even and odd subsets,
as explained in Sec.~5.)

Given a cohomology element $\chi_a\in H^2(K3,\IZ)$, the
hyperK\"ahler deformations of $K3$ that this element generates is

$$\d g_{mn} = \e^{a(i)}J_{(i)m}{}^p(\chi_a)_{pn}.$$
Here, $\e^{a(i)}$ is the deformation parameter and $J_{(i)}$ is the
triple of almost complex structures.  For example, for $\chi_I$, we
have $\e^{I(i)} = \d A^{Ii}$.  Under the $\IZ_2$ involution $\s_1$,
the almost complex structures $J_{(1)}$ and $J_{(2)}$ on the $K3$
are odd, while $J_{(3)}$ is even (cf.~App.~A).  So, with the choice
$J=J_{(3)}$, the $\IZ_2$ parities listed above tell us that the
$\IZ_2$ projection retains the moduli corresponding to the following
$\IZ_2$-even metric deformations:
\eqn\KahlerCpxEven{\eqalign{%
&\hbox{K\"ahler deformations generated by 
$\chi_3$, $\chi^3$ and $\chi_I$,}\cr
&\hbox{Complex structure deformations generated by
$\chi_{1,2}$ and $\chi^{1,2}$.}}}

Now, let us include the $T^2$ factor.  The geometric twists are
implemented by replacing the product $K3\times T^2$ with a nontrivial
fibration of $K3$ over $T^2$.  The topology of an arbitrary smooth
oriented fibration of $K3$ over $T^2$ is characterized by the two
group elements $\G_5(1),\G_6(1)\in\G_{3,19}$ under which the $K3$
moduli are identified as one takes $x^5\to x^5+1$, $x^6\to x^6+1$,
respectively.  For the parabolic subset of these fibrations discussed
in the previous section, the coset parameters $E,B,A$ of
Eqs.~\CosetRep\ and \SOvielbein\ simply shift through an integer
number of periods as we traverse the $x^5$ or $x^6$ circles.\foot{In
the case of the vielbein $E$, this means that the quantities $a^i{}_j$
that appear in the $T^3$ basis forms $E^i \propto dx^i+a^i{}_jdx^j$
for $i=1,2,3$ (cf.~the last subsection and App.~A) shift by integers.}
Let us focus on this subset of fibrations, for which the coset
parametrization \CosetRep\ is particularly adapted.  Then, for
example, for the blow-up modes, we write
\eqn\ModPlusTwist{A^{Ii} = A^{Ii}_0 + m^{Ii}_n x^n,
\quad n=5,6,\quad m^{Ii}_n\in2\IZ.}
The first component $A^{Ii}_0$ is the modulus component, the 0-mode on
$T^2$ with continuous deformations $\d A^{Ii}$. The second component
$m^{Ii}_n x^n$ depends explicitly on the $T^2$ coordinates and is the
discrete twist; it implements the even integer modular shifts
$\G_n(1)\colon\ A^{Ii}\to A^{Ii} + m^{Ii}_n$ for $x^n\to x^n+1$.

This type of generalized compactification, in which the moduli from
the first stage of the compactification (in this case, the $K3$) are
given dependence on the remaining compactification coordinates (in
this case, the $T^2$) in order to implement nontrivial twists by the
action of the modular group, is referred to as a Scherk-Schwarz
compactification~\refs{\ScherkZR,\KaloperYR}.  Generic Scherk-Schwarz
compactifications involves all of the moduli from first stage of the
compactification and not just the subset coming from the metric.
Hence, they are more general than geometric fibrations of this
section.  In Sec.~6.2, we will briefly mention nongeometric
Scherk-Schwarz compactifications based on the full $\G_{4,10}$ modular
group of type IIA on $K3$ instead of the $\G_{3,19}$ metric modular
group considered here.

Since we wish to implement geometric twists of the Calabi-Yau
$X=(K3\times T^2)/\IZ_2$, and not just of $K3\times T^2$, the moduli
and twists just discussed must be restricted to those that respect the
$\IZ_2$ isometry.  The moduli survive for the even $K3$ metric
deformations, as already described in \KahlerCpxEven.  In addition,
the discrete Scherk-Schwarz twists survive for the complementary set
of odd metric deformations, since these depend linearly on the odd
$T^2$ coordinates $x^n$.  Thus, from the blow-up modes of $K3$,
we obtain the following moduli and discrete geometric twists of $\TZZ
= (K3\times T^2)/\IZ_2$:
\eqn\ModuliTwists{\eqalign{%
\hbox{Moduli:}\quad & \d A^{I3},\cr 
\hbox{Twists:}\quad & A^{Ii} = m^{Ii}_nx^n,\quad i=1,2,\quad n=5,6,
\quad m^{Ii}_n\in4\IZ.}}
As should be clear by now, the integers $m^{Ii}_n$ partially
characterize the fibration of $K3$ over $T^2$.  Eq.~\ModuliTwists\
indicates which components are compatible with the $\IZ_2$
isometry. In addition, the subset of $\g^i_{nj}$ and $\b^{ij}_n$ that
respect the $\IZ_2$ isometry are those for which one of $i,j$ is equal
to $3$ and the other is equal to $1$ or $2$.  Comparing to Sec.~3.2,
the $m^{Ii}_n$ are genuinely new discrete twists of $\TZZ$ over those
inherited from the twisted $T^6$.  However, the $\b^{ij}_n$ are just
the $T^6$ twists in disguise.  This is most apparent in the orbifold
limit $K3=T^4/\IZ_2$.  As discussed in App.~A, the interpretation of the
modulus $B^{ij}$ is as the flat connection $a^4{}_i = -\half\e_{ijk}
B^{jk}$ which appears in 1-form $e^4 = R_4(dx^4 + a^4{}_j dx^j)$
needed to complete the $T^3(x^1,x^2,x^3)$ into a $T^4$.  Thus
$\g^4_{ni} = \half\e_{ijk}\b^{jk}_n$.  Finally, the integers
$\gamma^i_{nj}$, $\beta^{ij}_n$ and $\half m^{Ii}_n$ are required to
be even to to ensure the $\IZ_2$ quotient of the $K3$ fibration over
$T^2$ remains well defined as a fibration over $\IP^1=T^2/\IZ_2$.
This will become more transparent in Sec.~6.  There, we will see that
the monodromy about each of the four fixed points on $\IP^1$ is
$\G_5(\pm\half)\G_6(\pm\half)$, where the two $\pm$ signs are
uncorrelated.\foot{For example, in Sec.~6.2, we will see that $\ha
m^{Ii}_n$ is required to be even for the monodromies about individual
fixed points to lie in the modular group $\G_{3,19}$, which for our
parametrization of the $K3$ moduli space contains $A^{Ii}\to A^{Ii}+2$
but not $A^{Ii}\to A^{Ii}+1$.}  On $\IP^1$, the monodromies about the
$x^5$ and $x^6$ circles of $T^2$ becomes monodromies about {\it pairs}
of fixed points.

The twisted closure relation~\dzexplicit\ is expressed in terms of
$\zeta_a$ and $dx^n\w\zeta_a$, but can be re-expressed in terms of
the $\o_a$, $\a_A$ and $\b^A$ of Secs.~3.1 and 3.2.  The twist
matrices $M_a{}^A$ and $N_{aA}$ of Eq.~\dtwo\ that result from the
nontrivial K3 fibration described in this section are then found to be
\eqn\KthreeMN{M_a{}^A = \pmatrix{%
\g^2_{53} & 0 & \g^1_{63} & 0\cr
\noalign{\vskip2pt}
\b^{32}_5 & -\g^3_{61} & \b^{31}_6 & -\g^3_{52}\cr
0 & 0 & 0 & 0\cr
\noalign{\vskip6pt\hrule\vskip6pt}
0 & 0 & 0 & 0\cr
0 & 0 & 0 & 0\cr
m^{J2}_5 & 0 & m^{J1}_6 & 0},
\quad
N_{aA} = \pmatrix{%
0 & \g^1_{53} & 0 & -\g^2_{63}\cr
\noalign{\vskip2pt}
\g^3_{62} & \b^{31}_5 & \g^3_{51} & -\b^{32}_6\cr
0 & 0 & 0 & 0\cr
\noalign{\vskip6pt\hrule\vskip6pt}
0 & 0 & 0 & 0\cr
0 & 0 & 0 & 0\cr
0 & m^{J1}_5 & 0 & -m^{J2}_6}.}
Here, the $\omega_a$ run over $\o_\a$ and $\o_{\a I}$, for $\a=1,2,3$,
and there is an implicit $\d_{IJ}$ lowering the upper $J$ indices in
on the $m^{Ji}_n$.  As a check, note that the nonzero entries in the
first three rows agree with Eq.~\Mgamma.  The third row vanishes since
$\o_3$ is Poincar\'e dual to the generic $K3$ fiber in our
construction, which is boundaryless.  The fourth and fifth rows
vanish, since we incorporate only $\o_{3I}=\z_I$, the blow-up modes of
the $K3_{(3)}=T^4_{(3)}/\s_3$ fiber, in the twists and not the
remaining blow-up modes $\o_{1I}$ and $\o_{2I}$ of $\TZZ$.  When we
consider a type IIA orientifold based on this geometry in Sec.~4
below, we will require $M_a{}^A=0$.  In this case, note that an easy
way to satisfy the restriction to parabolic class, $(\g_n)^2 = 0$, is
to set $\g^3_{62}=\g^3_{51}=0$, so that the nonzero twists are
$N_{a1}$ and $N_{a3}$ for $a=1$, $2$ and $3I$.

\nonnumberedsubsec{Fibrations with nonlinear action on $K3$ moduli}

The matrix $M_p$ of Eq.~\Mexpression\ is not the most general Lie
algebra element of $\so(3,19)$.  More general monodromies are possible
that do not act linearly on the $K3$ moduli $E^\l{}_i$, $B^{ij}$ and
$A^{Ii}$.  An arbitrary element of $\so(3,19)$ can be written
\eqn\arbsoM{M_p =
\pmatrix{\g_p & \b_p & m^T_p\cr
-h_p & -\g^T_p & -n^T_p\cr
n_p & -m_p & -f_p}.}
Here, the new components compared to Eq.~\Mexpression\ are $n^I_{pi}$,
$h_{pij}$ (antisymmetric in $i,j$) and $f^{IJ}_p$ (antisymmetric in
$I,J$).  If we use the monodromy matrices $\G_p(x)$ constructed from
this general form for $M_p$ to define the global 2-forms $\z_a$
(cf.~Eq.~\zetachi), then the twisted closure relations \dzexplicit\
generalize to
\eqn\dzexplgen{\eqalign{%
d\z^i &= -dx^p\w(\g^i_{pj}\z^j +\b^{ij}_p\z_j + m^{Ii}_p\z_I),\cr
d\z_i & = dx^p\w(h_{pij}\z^j + \g^j_{pi}\z_j + n^I_{pj}\z_I),\cr
d\z_I & = dx^p\w\d_{IJ}(-n^J_{pi}\z^I + m^{Ji}_p\z_i + f^{JK}_p\z_K).}}
For $\TZZ = (K3_{(3)}\times T^2_{(3)})/\s_1$, the twist components
compatible with the $\IZ_2$ involution $\s_1$ are all of the
$n^I_{pi}$, the components $h_{p3i}$ for $i=1,2$, and none of the
$f^{JK}_p$.  As we will see in Sec.~5, for other Voisin-Borcea
manifolds, the $\IZ_2$ compatible twist components will change, and
some of the $f^{JK}_p$ will be retained as well.

When we include the additional data $n^I_{pi}$ and $h_{p3i}$, the
matrices $M_a{}^A$ and $N_{aA}$ of Eq.~\KthreeMN\ generalize to
\eqn\KthreeMNgen{M_a{}^A = \pmatrix{%
\g^2_{53} & -h_{631} & \g^1_{63} & -h_{532}\cr
\noalign{\vskip2pt}
\b^{32}_5 & -\g^3_{61} & \b^{31}_6 & -\g^3_{52}\cr
0 & 0 & 0 & 0\cr
\noalign{\vskip6pt\hrule\vskip6pt}
0 & 0 & 0 & 0\cr
0 & 0 & 0 & 0\cr
m^{J2}_5 & n^J_{61} & m^{J1}_6 & n^J_{52}},
\quad
N_{aA} = \pmatrix{%
h_{632} & \g^1_{53} & -h_{531} & -\g^2_{63}\cr
\noalign{\vskip2pt}
\g^3_{62} & \b^{31}_5 & \g^3_{51} & -\b^{32}_6\cr
0 & 0 & 0 & 0\cr
\noalign{\vskip6pt\hrule\vskip6pt}
0 & 0 & 0 & 0\cr
0 & 0 & 0 & 0\cr
-n^J_{62} & m^{J1}_5 & n^J_{51} & -m^{J2}_6}.}
Note that the nonzero entries in the first three rows agree with
Eq.~\Mgamma, provided that we make the identifications $\b^{ij}_p =
\e^{ijk}\g^4_{pk}$ (as discussed above) and $h_{pij} =
\e_{ijk}\g^k_{p4}$.

In summary, the result of the construction described here---using a
nontrivial $K3_{(3)}$ fibration to twist the Calabi-Yau manifold
$\TZZ$---is twist matrices $M_a{}^A$ and $N_{aA}$ that are nonzero for
$a = 1$, $2$ and $3I$, but vanishing for $a=3$, $1I$ and $2I$.


\newsec{The $\CN=1$ type IIA orientifold of $\TZZ$}

Let us consider the $\CN=1$ theory obtained from type IIA string
theory on $X=\TZZ$ via the orientifold operation described at the end
of Sec.~3.1.  This is the traditional setting for intersecting D6
brane model building, however, we will focus on the closed string
sector here.  In this section, our conventions and treatment are very
similar to those in Ref.~\AldazabalUP.  We neglect backreaction in the
form of warping and nontrivial dilaton profile.  This type of
backreaction is irrelevant for the purposes of studying moduli
stabilization in supersymmetric vacua.  To illustrate this, we restore
both the warping and dilaton profile in the first example below.

\subsec{Superpotential}

The perturbative superpotential is \refs{\DerendingerPH,\VilladoroCU,
\CamaraDC,\AldazabalUP,\GrimmUA,\BenmachicheDF}
\eqn\WNSRR{W = W_\NS + W_\RR,}
where
\eqn\superpot{\quad W_\NS = -\int_X \O_c\w(\bar H + dJ_c)
\quad\hbox{and}\quad W_\RR = \int_X e^{J_c}\w\bar F_\RR.}
Here, $\bar H$ and $\bar F_\RR$ denote the background, moduli
independent components of the flux only.  (This is the same notation
as that used in Ref.~\AldazabalUP.  For a discussion of the closure
properties of the various fluxes that appear in this paper, see
Sec.~4.3 below.)  The quantities $J_c$ and $\O_c$ are the complexified
K\"ahler form and 3-form appropriate to the orientifold, with lengths
measured in units of $(2\pi)^2\a'$,
\eqn\JcOc{\eqalign{J_c &= B+iJ = iT^a\o_a,\quad iT^a = b^a + iv^a,\cr
\O_c &= C_{(3)} + ie^{-\phi}\Re\O = iU^A\a_A.}}
Here, $B$ and $C_{(3)}$ are the continuous modulus components of the
NS 2-form and RR 3-form potentials, respectively.  They do not
contribute to the quantized background fluxes $\bar H$ and $\bar
F_\RR$.  Their components $B^a$ and $C^A_{(3)}$ with respect to the
basis forms $\o_a$ and $\a_A$ are independent of the internal
coordinates.  They contribute only continuous moduli dependent
contributions to the total flux, do to the lack of closure of the
latter:
\eqn\dBdC{\eqalign{dB &= B^a d\o_a = B^aN_{aA}\b^A,\cr
dC_{(3)} &= C^A_{(3)} d\b^A = -C^A_{(3)}N_{aA}\tilde\o^a.}}
Note that the $\b^A$ component of $C_{(3)}$ is removed by the
orientifold projection.

It is convenient to separate the K\"ahler moduli into $T^\a$,
$\a=1,2,3$, in the untwisted sector (cf.~Eq.~\twofourforms) and $T^{\a
I}$, $\a=1,2,3$, $I = 1,\ldots,16$, in the twisted sector
(cf.~Eq.~\Xsecondcohom).  Likewise, the moduli $U^A$ separate into
complex structure moduli $U^\a$ and the 4D dilaton-axion $U^0=S$.  Our
conventions for $J$, $\O$ and the volume form are
\eqn\JOvol{{i\over8}\O\w\bar\O = {1\over6}J\w J\w J = \Vol_X.}

The complex structure moduli come solely from the untwisted sector of
the orbifold.  Their geometric components are are simply the $T^2$
complex moduli $\t_a$, which by the orientifold projection, are
required to be purely imaginary, $\t_\a = it_\a$.  In light of the
normalization convention \JOvol, the $(3,0)$ form
\eqn\Ohat{\hat\O = 2(dx^1 + it_1 dx^2)\w (dx^3 + it_2 dx^4)\w
(dx^5 + it_3 dx^6)}
is related to $\O$ via
\eqn\OOhat{\O = \Bigl({V_X\over t_1t_2t_3}\Bigr)^{1/2} \hat\O,}
where $V_X = \int\Vol_X$ is the volume of $X$.  In terms of the $\a_A$
and $\b^A$, we have
\eqn\ReImOhat{\Re\hat\O = \a_0+\half\e^{\a\b\g}t_\a t_\b\a_\g
\quad\hbox{and}\quad\Im\hat\O = t_\a\b^\a + t_1t_2t_3\b^0.}
The apparent K\"ahler modulus dependence of $\O_c$ through the volume
dependence \OOhat\ is an artifact of expressing $\O_c$ in terms of the
10D rather than 4D dilaton.  In terms of the 4D T-duality invariant
dilaton $e^{\phi_4} = e^\phi/\sqrt{\mathstrut V_X}$, we have
\eqn\Ocompensator{\O_c = C_{(3)} + i\Re(C\hat\O),}
where $C$ is the compensator field
\eqn\compensator{C^{-1} = e^{\phi_4}
\Bigl(\iover8\int\hat\O\w\hat{\bar\O}\Bigr)^{1/2}
= e^{\phi_4}\bigl(t_1t_2t_3)^{1/2}.}

The first term in the superpotential comes from NS sector discrete
data (NS flux and geometric twists).  To evaluate this term, let us
write
\eqn\Hcomponents{\bar H = \bar H^A\a_A - \bar H_A\b^A,\quad\hbox{with}\quad
\bar H^A, \bar H_A \in2\IZ.}
For the orientifold, both $\bar H^A$ and the geometric twists $M_a{}^A$ of
Sec.~3.1 are projected out.  Thus, $dJ_c = iT^ad\o_a = iT^a
N_{aA}\b^A$, and we obtain
\eqn\evalWNS{W_\NS(T,U) = iU^A (\bar H_A - iT^a N_{aA}).}
Note that the condition $M_a{}^A=0$ implies that $d\bigl(\Im\O\bigr) =
0$, so that the geometry is what called {\it half flat}.\foot{When the
backreaction on the geometry in the form of nontrivial warping and
dilaton profile are included, this condition becomes
$d(e^{-\phi/3}\Im\O) = 0$, so that the geometry is instead conformally
half flat.}

The second term in the superpotential comes from RR sector discrete
data (RR flux).  In this case, we write
\eqn\Fcomponents{\bar F_{(0)} = q^0,\quad \bar F_{(2)} = q^a\o_a,\quad 
\bar F_{(4)} = p_a\tilde\o^a,\quad \bar F_{(6)} = p_0\a_0\w\b^0,}
where\foot{We take the integer quantized NS and RR flux components to
be even in order to avoid subtleties involving exotic orientifold
planes.}  $q^0,q^a,p_a,p_0\in2\IZ$.  Then,
\eqn\evalWRR{W_\RR(T) = -q^0\CF(T) -q^a\CF_a(T) + p_a iT^a + p_0,}
where $\CF(T)$ is the quantum volume of $X$, with large radius
expression
\eqn\Kprepot{-\CF(T) = {1\over6}\int J_c\w J_c\w J_c 
= {1\over6}\k_{abc}iT^a iT^b iT^c,}
and $\CF_a = \partial\CF/\partial(iT^a)$. The intersection numbers
$\k_{abc} = \int_X\o_a\w\o_b\w\o_c$ can be found in App.~B.  

At finite radius, $\CF(T)$ receives corrections from worldsheet
instantons.  Note that $\CF(T)$ is not quite the same as the
prepotential for the K\"ahler moduli in the parent $\CN=2$ theory, due
to corrections from unoriented worldsheets.  The former has been
computed for the Voisin-Borcea class (along with the topological
string amplitudes for higher genus) \GrimmTM, but the latter are
still unknown.

In addition to worldsheet instanton corrections, the superpotential
receives \hbox{D-instanton} corrections from Euclidean D2~branes
wrapping generalized special Lagrangian 3-cycles \BlumenhagenXT.  The
1-instanton contribution from a D2~brane wrapping the 3-cycle $C_A
A^A$ takes the form $\hbox{Pfaff}(T)e^{-2\pi C_AU^A}$, where
$\hbox{Pfaff}(T)$ is the 1-loop Pfaffian prefactor.  These are the
mirrors of the D3~instantons in KKLT \KachruAW\ type IIB backgrounds.
Similarly, from gaugino condensation on stacks of D6 branes wrapping
the 3-cycle $C_A A^A$ one can obtain corrections of fractional
D2~instanton number.  This is the mirror of gaugino condensation on D7
brane stacks in type IIB.

\subsec{K\"ahler potential}

The K\"ahler potential is $K = K_1(T,\bar T) + K_2(U,\bar U)$, where
at large radius \refs{\AldazabalUP,\GrimmUA,\BenmachicheDF}
\eqn\Kahlerpot{\eqalign{K_1(T,\bar T) 
&=-\log{4\over3}\int_X J\w J\w J
= -\log{1\over6}
\k_{abc}(T^a+\bar T^a)(T^b+\bar T^b)(T^c+\bar T^c),\cr
K_2(U,\bar U) &= -2\log{i\over2}\int_X C\hat\O\w \bar C\hat{\bar\O}
= - \sum_{\a=0}^3\log(U^\a+\bar U^\a),}} 
for the K\"ahler moduli and the combined dilaton-axion/complex
structure moduli, respectively.  At finite radius, the K\"ahler
potential receives corrections from both worldsheet instantons and
D-brane instantons.

\subsec{Bianchi identities}

The only nontrivial Bianchi identity (tadpole cancellation condition)
for the background flux is that for $\bar F_{(2)}$:
\eqn\FtwoBianchi{d\bar F_{(2)} = \bar H\w \bar F_{(0)} + j_{\rm D6,O6},}
where $j_{\rm D6,O6}$ is the source term due to D6 branes and O6
planes.

Let us write
\eqn\HHbarFFbar{H = \bar H + dB,\quad F_{(4)} = \bar F_{(4)} + dC_{(3)},}
and $\bar F_{(n)} = F_{(n)}$ for $n=0,2,6$.  In addition, let us
define $\tilde F_\RR = e^B F_\RR$.  Note that, in contrast to
Eq.~\FtwoBianchi, the $\tilde F$ satisfy the modified Bianchi
identities with total (background plus moduli dependent) flux,
\eqn\FtBianchi{d\tilde F_{(8-p)} = H\w F_{(6-p)}
+ j_{{\rm D}p,{\rm O}p}.}
The conditions \FtBianchi\ constrain the moduli $B$ and $C_{(3)}$,
however we need not concern ourselves with these non-topological
constraints here.  These constraints are automatically taken care of
by the supersymmetry conditions below.  For later reference, the
tilded fluxes appearing in the supersymmetry conditions are
\eqn\tfluxes{\eqalign{\tilde F_{(2)} &= \bar F_{(2)}+B\w \bar F_{(0)},\cr
\tilde F_{(4)} &= \bar F_{(4)} + dC_{(3)} + B\w F_{(2)} + \ha
B\w B \bar F_{(0)},}}
and $\tilde F_{(6)}$, whose precise decomposition we will not need.

Now let us return to the topological Bianchi identity \FtwoBianchi.
Using Eq.~\Fcomponents, together with the relations
\eqn\ABab{\eqalign{A^B\cap B_A = \int_{A^B}\a_A 
&= \int_X\a_A\w\b^B = \d_A{}^B,\cr
B_A\cap A^B = \int_{B_A}\b^B &= \int_X\b^B\w\a_A = -\d_A{}^B}}
(i.e., $A^B,B_A$ Poincar\'e dual to $\b^B,\a_A$, respectively), we obtain
\eqn\IntBianchi{-(q^0\bar H_A+q^aN_{aA})A^A + \sum_\n N^\n(\pi_\n+\pi'_\n)
=4\pi_{\rm O6}.}
Here, the sum runs over stacks of $N^\n$ D6 branes wrapping homology
class $\pi_\n$ and $N^\n$ image D6 branes wrapping homology class
$\pi'_\n$.  Writing, as in Ref.~\BlumenhagenXX,
\eqn\pidefs{\eqalign{\pi_\n &= \sum_{A=0}^3(X_{\n A}A^A+Y_\n^AB_A),\cr
\pi'_\n &= \sum_{A=0}^3(X_{\n A}A^A-Y_\n^AB_A),\cr
\pi_{\rm O6} &= 4\sum_{A=0}^3 A^A,}}
the Bianchi identities become\foot{The change in sign of the relative
contribution to $\pi_{\rm O6}$ from $A = 0$ and $A=1,2,3$ as compared
to Refs.~\AldazabalUP\ and \BlumenhagenXX, comes from the different
convention used in this paper for the $\alpha_A,\beta^A$.  In the
absence of geometric twists, these Bianchi identities were first
derived for $\TZZ$ in Ref.~\CveticTJ.}
\eqn\IntBianchiSimp{-\half (q^0\bar H_A+q^aN_{aA}) + \sum_\n N^\n 
X_{\n A} =8,\quad\hbox{for}\quad A=0,1,2,3.}

In the next section, we will discuss supersymmetric Minkowski vacua,
for which $q^0 = 0$ and $d\bigl(e^{-\phi}\Re\O\bigr) = \star\bar
F_{(2)}$.  In this case,
\eqn\qNpositivity{0\le \int_X \bar F_{(2)}\w \star\bar F_{(2)}
= \int_X \bar F_{(2)}\w d\bigl(e^{-\phi}\Re\O\bigr) = -(q^aN_{aA})\Re
U^A.}
The geometric regime is $\Re U^A>0$.  In the case that the solutions
have a moduli space in which all of the $\Re U^A$ can be varied
independently, it follows that the contribution $-\half (q^0\bar
H_A+q^aN_{aA})$ to the Bianchi identity from discrete data is
nonnegative.  We suspect that this is true in general for
supersymmetric Minkowski vacua, irrespective of the resulting moduli
space.  (For AdS vacua, see Ref.~\CamaraDC.)

\subsec{Supersymmetric Minkowski vacua}

The scalar potential from $F$-terms in $\CN=1$ supergravity is
\eqn\scalarpot{\CV(\Phi,\bar\Phi)
= e^K\Bigl(\sum_{P,Q}K^{P\bar Q} D_P W(\Phi)
\bar D_{\bar Q} \bar W(\bar\Phi)
- 3\bigl|W(\Phi)\bigr|^2\Bigr),}
where $D_P = \partial_{\Phi^P} -K_{,\Phi^P}$ for each modulus
$\Phi^P$.  Supersymmetric vacua satisfy $D_P W = 0$ for all $\Phi^P$.
Thus, $V = -3e^K\bigl|W\bigr|^2$, which generically gives Anti de
Sitter vacua.  However, in the special case $W=0$, we obtain Minkowski
vacua.  Let us focus on this case.  Then the supersymmetry conditions
become considerably more manageable, in that they become truly
holomorphic equations $\partial_{\Phi^P} W(\Phi) = 0$, as opposed to
just covariantly holomorphic.

\nonnumberedsubsec{The holomorphic monopole equations}

From the superpotential \WNSRR, the supersymmetry conditions that we
obtain in this way are
\eqn\SUSYeqs{d\O_c +(e^{J_c}\w\bar F_\RR)_{(4)} = 0,
\quad\bar H + dJ_c = 0.}
The imaginary and real parts are 
\eqn\SUSYeqsIm{d\bigl(e^{-\phi}\Re\O\bigr)+J\w\tilde F_{(2)} = 0,
\quad dJ = 0,}
and
\eqn\SUSYeqsRe{\tilde F_{(4)} - \ha J\w J\bar F_{(0)} = 0,
\quad H = 0,}
respectively.  In addition, there is the Minkowski condition $W=0$.
When combined with Eq.~\SUSYeqsRe\ and \SUSYeqsIm, the Minkowski
condition implies that the $SU(3)$ singlet components of the tilded
fluxes vanish,\foot{This involves observing that the same $SU(3)$
singlet torsion component $W_1$ appears in both $dJ =
-\smallfrac{3}{2}\Im(W_1\bar\O) + W_4\w J + W_3$ and $d\O = W_1J^2 +
W_2\w J + \bar W_5\w\O$, and that $W_1 = 0$ from the second equation
in \SUSYeqsIm.}
\eqn\SingletEq{\tilde F_{(6)} = J\w\tilde F_{(4)} 
= J\w J\w\tilde F_{(2)} = J\w J\w J F_{(0)} = 0.}
Thus, the supersymmetry conditions reduce to
\eqn\HoloMonopole{d\bigl(e^{-\phi}\Re\O\bigl) 
+ J\w\bar F_{(2)} = 0,\quad dJ=0,
\quad\hbox{$\bar F_{(2)}$ primitive,}}
with $\tilde F_{(6)} = \tilde F_{(4)} = F_{(0)} = 0$.  Here, $\bar
F_{(2)}$ primitive means that $\bar F_{(2)}\w J\w J = 0$.  These are
the {\it holomorphic monopole} equations discussed in
Refs.~\refs{\KasteXS,\GranaBG}.  As already mentioned in a footnote
above, one can also show that the imaginary part of $\O$ satisfies
$d(e^{-\phi/3}\Im\O) = 0$.

Note that the first condition in Eq.~\HoloMonopole\ is a generalized
calibration condition.  Since $\star\bar F_{(2)} = - J\w\bar F_{(2)}$
for $\bar F_{(2)}$ primitive, it can be rewritten as
\eqn\GenCal{d\bigl(e^{-\phi}\Re\O\bigl) = \star\bar F_{(2)}.}
The equation relates the generalized calibration $e^{-\phi}\Re\O$,
which calibrates (serves as the volume form on) the generalized
special Lagrangian cycles on which we can wrap D6 branes, to the flux
$\bar F_{(2)}$, which is sourced by D6 branes.

The holomorphic monopole background is the Kaluza-Klein reduction of
M~theory on $\IR^{3,1}$ times a manifold of $G_2$ holonomy, using the
the M~theory circle for the reduction.  The lift to a M~theory on
$G_2$ is guaranteed on general grounds by $\CN=1$ supersymmetry and
the fact that the $\tilde F_{(2)}$ flux, D6 branes and O6 planes each
have purely geometrical M~theory lifts.  Due to the O6~planes in the
IIA background, the M~theory geometry does not actually have a circle
fiber with a $U(1)$ isometry on which to dimensionally
reduce.\foot{Indeed, there is a theorem that a compact manifolds of
special holonomy cannot have continuous isometries.}  As discussed in
Refs.~\refs{\SchulzUB,\SchulzTT}, the reduction involves first
truncating to the lowest Kaluza-Klein modes of an approximate $U(1)$
isometry that fails only locally near the regions of the M~theory
geometry that become the O6 planes in IIA.  The neglected M~theory
corrections from higher Kaluza-Klein modes are corrections from D0
bound states in IIA.  These corrections are appreciable only near the
O6 planes, where the string coupling diverges and the D0 bound states
become light.

Although the conclusion is that we have rediscovered the class of
M~theory compactifications on $G_2$ holonomy manifolds, the
description in terms of the $\TZZ$ orientifold with well defined
discrete data (D6 branes, $\bar F_{(2)}$ flux and geometric twists)
should provide a useful route to studying these $\CN=1$
compactifications for model building purposes that is more explicit
than that which currently exists from the $G_2$ perspective.

\nonnumberedsubsec{Simplifying assumptions}

In light of the holomorphic monopole equations obtained in the
previous section, we set $\bar H =\bar F_{(6)} = \bar F_{(4)} = 0$ for
simplicity from now on.  For choices of discrete data that are
compatible with the supersymmetry conditions, giving nonzero values to
these background fluxes just leads to compensatingly shifted
expectation values for some of the axionic moduli, to ensure that the
corresponding tilded fluxes vanish, as required by the supersymmetry
conditions.  This simplifying assumption can be thought of a discrete
gauge choice. Note that since $q^0\equiv\bar F_{(0)} = 0$ for
Minkowski vacua, $\bar H_A$ disappears from the quantity
$-\half(q^0\bar H_A+q^a N_{aA})$ that appears in the RR tadpole
cancellation conditions.

\nonnumberedsubsec{General observations and validity of the classical
supergravity analysis}

Classically, the conditions $\partial_P W = 0$ and $W = 0$ are
homogeneous of degree 1 and 2 respectively under rescaling of all
moduli $(T^a,U^A)\to (\l T^a,\l U^A)$.  This means that for generic
choice of discrete data $q^a,N_{aA}$, and neglecting worldsheet and
D2~instanton corrections, the moduli $T^a$ and $U^A$ all run away to
zero.  In this case, the instanton corrections will be critical in
understanding the stabilization of moduli.  On the other hand, for
nongeneric discrete data, we can classically have a nontrivial moduli
space in which ratios of moduli are fixed but the overall scale is
not.  Recall that the complex structure moduli are actually the
inhomogeneous coordinates $U^A/U^0$, where $U^0=S$ is the
dilaton-axion.  Fixing ratios of $T^a,U^A$ but not the overall scale
means that we can fix complex structure moduli and relative K\"ahler
moduli and/or their ratios, as well as the ratio of the overall volume
modulus to the dilaton-axion $S$.  However, the dilaton-axion is left
freely tunable.  At large $S$, we are in the weak coupling regime and
the large volume regime for the nonzero K\"ahler moduli.  On can then
hope to simultaneously break supersymmetry and lift the dilaton and any
remaining complex structure moduli by including anti-D6 branes and D2
instantons (and/or gaugino condensation), in a construction mirror to
that of KKLT \KachruAW.

Note that if any of the K\"ahler moduli vanishes, then we are clearly
in a regime where the large volume classical supergravity description
is not valid and worldsheet instanton corrections are critical.  We
will not have much to say quantitatively about this regime here.
However, note that for orbifolds, the extreme opposite of the large
volume regime is also computationally accessible in the following
sense.  When all moduli in the orbifold untwisted sector are large,
there is a perturbative expansion about the orbifold limit, in powers
of the small blow-up moduli.  (See, for example, Ref.~\CveticAC.)
This will be relevant for putting our examples of Minkowski vacua on
firmer footing.  In three of the examples we present below, we will
find that that we are in exactly this situation---in the classical
supergravity analysis, all K\"ahler moduli in the orbifold untwisted
sector of $(K3\times T^2)/\IZ_2$ can be taken large, while those in
the orbifold twisted sector are required to vanish.  We leave the
requisite conformal field theory analysis about the orbifold limit for
future work.

Finally, note that worldsheet instanton corrections are also important
when the K\"ahler moduli are of order $\alpha'$.  We will never be
forced to this regime for the Minkowski vacua discussed here, due to
the homogeneity property.  However, it is generically the case for the
vacua discussed in Ref.~\AldazabalUP.  There, nongeometric twists are
included as well, so the superpotential is no longer homogeneous, and
the supersymmetry condition become polynomial equations whose roots are
generically of order~1 in $\a'$ units.  Therefore, the analysis of
Ref.~\AldazabalUP\ is technically only applicable when viewed as the
truncation of the $\CN=4$ orientifold on $T^6$ to the diagonal
$T^2\times T^2\times T^2$.  In this case, the degree of supersymmetry
protects the theory against worldsheet instantons, but the
$\IZ_2\times\IZ_2$ orbifold is subject to order 1 corrections.

\nonnumberedsubsec{Example 1: Twists inherited from $T^6$ only.}

This example includes $\bar F_{(2)}$ flux and geometric twists in the
orbifold untwisted sector of $\TZZ$ only.  It illustrates two
important features of the Minkowski vacua of the IIA orientifold.
First, when the blow-up moduli are included in the analysis, they can
nonetheless still be fixed due to the interplay of twisted and
untwisted sector in the intersection numbers of $\TZZ$, which enter
into the superpotential contribution $W_\RR(T,U)$.  Second, we claimed
above that the backreaction (warping) of the geometry and nontrivial
dilaton due to the presence of the D6~branes and O6~planes, were
irrelevant for the purposes of studying moduli stabilization in
supersymmetric vacua.  In this example, we restore the warp factors
$Z$ and dilaton profile to illustrate this point.

For simplicity, let us solve the Bianchi identities
\IntBianchiSimp\ for $A=0,1,2$, by locally cancelling the O6 charge
with coincident D6 branes.  For $A=3$, we include $N=8-4qn$ individual
$D6$ branes at arbitrary positions $x_\n^1,x_\n^3,x_\n^6$ for
$\n=1,\ldots,8-4qn$.  (The integers $q$ and $n$ will be defined
below.)

On the $\IR^{3,1}\times \hbox{twisted-}T^6$ covering space, the geometry is
\eqn\ExMetric{ds^2 = Z^{-1/2}(\eta_{\m\n}dx^\m dx^\n
+ ds^2_{\parallel})+Z^{1/2}ds^2_{\perp}}
where
\eqn\ExTmetrics{\eqalign{%
ds^2_{\parallel} &= (R_2\eta^2)^2 + (R_4\eta^4)^2 + (R_5 dx^5)^2,\cr
ds^2_{\perp} &= (R_1 dx^1)^2 + (R_3 dx^3)^2 + (R_6 dx^6)^2,}}
are the metrics on the $T^3$ factors parallel and perpendicular to the
worldvolumes of the D6~branes that wrap cycles of homology class
$A^3$.  The $(1,0)$-forms are spanned by the complex vielbein
\eqn\ExIeta{\eqalign{%
\eta^{z^1} &= Z^{1/4}R_1dx^1 +iZ^{-1/4}R_2\eta^2,\cr
\eta^{z^2} &= Z^{1/4}R_3dx^3 +iZ^{-1/4}R_4\eta^4,\cr
\eta^{z^3} &= Z^{-1/4}R_5dx^5 +iZ^{1/4}R_6 dx^6.}}
Then, on $\TZZ$, if we set the blow-up moduli to zero, the fundamental
form\foot{When the fundamental form $J$ is closed it is called the
K\"ahler form.} and $(3,0)$ form are
\eqn\ExJO{\eqalign{%
J &= 2v^1dx^1\w\eta^2 + 2v^2 dx^3\w\eta^4 + 2v^3 dx^5\w dx^6,\cr
\O &= 2\eta^{z^1}\w\eta^{z^2}\w\eta^{z^3},}}
where $v_1 = R_1 R_2$, $v_2 = R_3 R_4$ and $v_3 = R_5 R_6$.

Let us choose $T^6$ twists
\eqn\ExTTtwists{d\eta^2 = - 2n dx^3\w dx^6\quad\hbox{and}\quad
d\eta^4 = -2n dx^1\w dx^6.}
This gives twist data $\g^2_{63} = -2n$ and $\g^4_{61}
\ (=-\b^{32}_6) = -2n$, so that Eq.~\Mgamma\ becomes
\eqn\ExNmatrix{N_{aA} = 
\pmatrix{0 & 0 & 0 & 2n\cr
0 & 0 & 0 & -2n\cr 0 & 0 & 0 & 0}.}

In the RR sector, we take
\eqn\ExRRflux{\bar F_{(2)} = g_s^{-1}\star_3dZ - 4q \bigl(dx^1\w\eta^2
 - dx^3\w\eta^4).}
Here, the first term is the backreaction from the D6 branes and O6
planes, with $\star_3$ the Hodge star operator in the metric
$ds^2_\perp$.  The second term gives the discrete RR data $q^1=-2q$ and
$q^2 = 2q$.  

The dilaton acquires a nontrivial profile in the extra dimension.  We
have $e^\phi = g_s Z^{-3/4}$.  The warp factor satisfies
\eqn\warpfactor{-\nabla^2_\perp Z = g_s
\bigl(N^{\rm discrete}_3 + \sum_{\rm D6,O6} Q_{\rm D6,O6}
\d^3(x-x_{\rm D6,O6})
\bigr)/\sqrt{g_\perp},}
where 
\eqn\Ndiscrete{N^{\rm discrete}_A = -2(q^0\bar H_A + q^aN_{aA}).}
Here, $Q_{\rm D6} = 1$ and $Q_{\rm O6} = -4$.  The sum runs over all
O6~planes, D6~branes and image D6~branes that wrap cycles in $T^6$
which project to the class $A^3$ in $\TZZ$.  On the $T^6$ covering
space (with respect to both the orientifold and the $\IZ_2\times\IZ_2$
orbifold), each of these D6~branes appears as part of a quadruple,
consisting of one D6~brane plus three image D6~branes.\foot{This is
somewhat counterintuitive.  One might have expected the the
$\IZ_2\times\IZ_2$ orbifold introduces a factor of 4 in the number of
D6 branes plus images, and then the orientifold introduces another
factor of 2, for a factor of 8 rather than 4 total.  This is not the
way it works.  Since the D6~branes wrap SLAG cycles, while the
orbifold $\IZ_2\times\IZ_2$ inverts holomorphic cycles, one finds that
orbifold doubles rather than quadruples the number of branes needed on
the covering space.  For $2N$ parallel D6~branes on the covering
spaces, one finds $U(N)\to U(N/2)\times U(N/2)\to U(N/2)$, upon
implementing the first and then second orbifold $\IZ_2$ projection.
So indeed, the rank decreases by a factor of 2 rather than 4 from the
orbifold $\IZ_2\times\IZ_2$.}  In addition, there are $8$ parallel O6
planes, located at the points on the transverse space
$T^3_\perp(x^1,x^3,x^6)$ where each coordinate is equal to $0$ or
$1/2$.  The total source charge on the right hand side of
Eq.~\warpfactor\ is zero, as required in order to solve Poisson's
equation on a compact manifold.  The contributions are $N^{\rm
discrete}_3 = 16qn$, $\sum Q_{\rm O6} = -32$ from eight O6 planes, and
$\sum Q_{\rm D6} = 4(8-4qn)$ from $(8-4qn)$ quadruples of D6~branes
plus images.

One can check that this background indeed satisfies the holomorphic
monopole equations, provided that the moduli satisfy the constraints
$T^1 = T^2$ and $U^3 = (2q/n)T^3$ derived below.  In particular, the
equations
$$d(e^{-\phi}\Re\O) + J\w\bar F_{(2)} = 0,
\quad d(e^{-\phi/3}\Im\O)=0,\quad dJ=0,$$
are satisfied, even when one includes the nontrivial dilaton profile
and warping of the geometry.

We now derive the moduli constraints using the superpotential $W =
W_\NS+W_\RR$.  From the matrix $N_{aA}$ above, the NS
superpotential is
\eqn\ExWNS{W_\NS = 2nU^3(T^1-T^2).}
The RR superpotential is
\eqn\ExWRRprelim{W_\RR = 2q \CF_1 - 2q\CF_2.}
Let us first neglect the blow-up modes.  Then,
\eqn\ExWRRnoblowup{\eqalign{W_\RR 
&= \Bigl(2q{\partial\over\partial(iT^1)}
-2q{\partial\over\partial(iT^2)}\Bigr)(-2iT^1iT^2iT^3)\cr
&= -4q(T^1-T^2)T^3.}}
So, the total superpotential is
\eqn\ExTotalW{W = W_\NS+W_\RR = (2nU^3-4qT^3)(T^1-T^2).}
The supersymmetry conditions are $T^1 = T^2$ from $\partial_T W = 0$
and $U^3 = (2q/n)T^3$ from $\partial_U W = 0$.  

We can also obtain these constraints by duality as follows. This
background (without the $\IZ_2\times\IZ_2$ orbifold and D6/O6 stacks
for $A=0,1,2$) was studied in Ref.~\SchulzTT\ as in intermediate
background in the duality chain between the type IIB orientifold
$T^6/\O(-1)^{F_L}\CI_6$ \KachruHE\ and a type IIA Calabi-Yau dual.  In
the IIB orientifold, the first constraint is the condition
$\det_{2\times2}\tau = -1$ on a $T^4$ factor and the second is
$\tau_3\tau_{\rm dil} = -1$ on a $T^2$ factor.  These are the type IIB
supersymmetry conditions that the complex 3-form flux be (2,1) and
primitive.\foot{For comparison to Ref.~\SchulzTT, note that $2q$ here
equals $m$ there, and that the $(1,2,3,4,5,6)$ directions here are the
$(6,5',7,4',9',-8)$ directions there.  For the moduli constraints, the
correspondence is that $T^1 = T^2$, $(2q/n)T^3 = U^3$ here is $\t'_1 =
\t_2$, $g'_sR_8 = (n/m)v'_1$ of Ref.~\SchulzTT.}

If we include the blow-up modes, then using the classical intersection
matrix for $\TZZ$ given in App.~B, the new RR superpotential is
\eqn\ExWRR{W_\RR = 4q\bigl(T^2T^3 - \sum_I(T^{1I})^2\bigr)
-4q\bigl(T^1T^3 - \sum_I(T^{2I})^2\bigr).}
Varying with respect to the blow-up K\"ahler moduli $T^{\a I}$, we
obtain $T^{1I}=T^{2I}=0$ and $T^{3I}$ arbitrary.  Then, varying with respect
to $T^{\a}$ and $U^A$ gives the same constraints as before.

This example shows that even when we have not included twists or RR
data that involve the blow-up modes of $\TZZ$, the blow-up moduli can
nevertheless be stabilized due to the interplay of orbifold twisted
sector and untwisted sector in the intersection form.

Of course, our large radius analysis is not reliable at
$T^{1I}=T^{2I}=0$ and we should instead use the appropriate $\CF(T)$
for a very small resolution of the orbifold.  However, all is not
lost, since the latter complementary regime is also computationally
accessible from the orbifold conformal field theory for $(K3\times
T^2)/\IZ_2$, as already mentioned above.  We will not need the full
computation of $\CF(T)$, but can make do with a pair of results
analogous to those proven in Ref.~\CveticAC\ in the slightly different
context of the heterotic $\IZ_3$ orbifold.  Assuming that these
results carry over here as well, we would find the following:
$\partial^2\CF/\partial T^{\rm ut}\partial T^{\rm tw}$ vanishes at
$T^{\rm tw}=0$ and $\partial^2\CF/\partial T^{\rm ut}\partial T^{\rm
ut}$ gives the classical result at $T^{\rm tw}=0$ and large $T^{\rm
ut}$, where $T^{\rm ut}$ and $T^{\rm tw}$ are the orbifold untwisted
and twisted sector K\"ahler moduli, respectively.  This would imply
that the superpotential extrema obtained above at $T^{1I}=T^{2I}=0$
persist in the presence of corrections from worldsheet instantons.
For now we leave the problem open, but we hope to return to the
question of the validity of this result in future work.

\nonnumberedsubsec{Example 2: The shrinking $K3$ surface}

The additional discrete data at our disposal that was not used in the
previous example is the blow-up mode twist data $N_{A,3I}$, and the RR
flux data $q^3$ and $q^{3I}$.  The real complication of the
supersymmetry conditions comes from the intersection form of $\TZZ$,
which appears in $W_\RR$.  Therefore, as a first step toward
generalization, let us attempt to find supersymmetric Minkowski vacua
in which the $q^{3I}$ still vanish.  The complete solution can be
described as follows.

From $\partial W/\partial T^\a=0$, we have
\eqn\TfromABC{T^1 = {1\over2q^2q^3}(A-B),\quad
T^2 = -{1\over2q^1q^3}(A-B),\quad
T^3 = -{1\over2q^1q^2}(A+B),}
where $A = -\half q^1N_{1A}U^A$ and $B=-\half q^2N_{2A}U^A$.  For the
geometric regime $\Re T^\a >0$, we take $q^1$ negative, $q^2,q^3$
positive, and require that $A>B>0$.  From $\partial W/\partial T^{\a
I}=0$, we find
\eqn\ExtwoTaI{T^{1I}=T^{2I}=0\quad\hbox{and}\quad T^{3I}=
-{1\over2q^3}N_{3I,A}U^A.}
So, we are again forced to the $(K3\times T^2)/\IZ_2$ orbifold limit.
Finally, from $\partial W/\partial U^A = 0$, we have $\Pi_{AB}U^B =
0$, where
\eqn\Uprojector{\Pi_{AB} = {1\over4|q^1q^2q^3|}
\Bigl( (q^1N_{1A}-q^2N_{2A})(q^1N_{1B}-q^2N_{2B})
- 2|q^1q^2|\sum_I N_{3I,A}N_{3I,B}\Bigr).}
For compatibility of this last condition with the geometric regime
$\Re U^A>0$ and $A>B>0$, the $N_{aA}$ must be chosen in such a way
that $\Pi_{AB}=0$. The condition $W=0$ is then automatically
satisfied.

However, note the following pathology of this solution.  The volume of
the $K3$ fiber, $V(K3_{(3)}) = 2T^1T^2-\sum_I(T^{3I})^2$, vanishes.
This can be seen explicitly from Eqs.~\TfromABC\ through \Uprojector\
above.  It can also be seen more directly as follows.  For any
supersymmetric Minkowski vacuum, we can show that not only does $W$
vanish, but $W_\RR$ and $W_\NS$ each vanishes separately.  In the
present example, $\partial W/\partial T^3 = 0$ gives $2q^1T^2+2q^2T^1
= 0$.  Using only this equation, $W_\RR = -q^3V(K3_{(3)})$.

This is roughly analogous to the result that for type IIB vacua,
including 3-form flux through an $A$-cycle, and no flux through the
corresponding $B$-cycle, forces the geometry to a conifold point in
which the $B$-cycle shrinks to zero size.  In our example, we have
chosen RR flux $q^3$ through the 2-cycle Poincar\'e dual to
$\tilde\o^3$, without geometric flux $d(e^{-\phi}\Re\O) =
-N_{aA}U^A\tilde\o^a$ through the dual 4-cycle $K3_{(3)}$ (which is
Poincar\'e dual to $\o_3$).  In another context, this would perhaps
lead to a Calabi-Yau singularity in which a 4-cycle locally shrinks
(to i.e., a del Pezzo singularity), but here that cannot happen, so
the whole $K3$ fiber shrinks.  Of course, this example is well outside
the regime of validity of the classical supergravity.  It was provided
for heuristic purposes only.

\nonnumberedsubsec{Example 3: Twists involving $K3$ blow-up modes}

Given the lesson learned in the previous example, we can hope to find
a supersymmetric Minkowski vacuum without shrinking divisors, by
permitting only $q^1,q^2,q^{3I}$ and the dual topological data
$N_{1A},N_{2A},N_{3I,A}$ to be nonzero.  For simplicity, we take the
blow-up mode data to be nonzero only for $a=3I|_{I=1}$, and for
notational convenience, write $a=4$ to denote $a=3I|_{I=1}$.  In order
to remain in the parabolic case discussed in Sec.~3.3, we set
$N_{aA}=0$ unless $A=1,3$.

From $\partial_a W =0$ together with $W_\RR=0$, we find that the
$(K3\times T^2)/\IZ_2$ blow-up moduli vanish ($T^{1I}=T^{2I}=0$), all
of the $T^{3I}$ are arbitrary, and $T^1,T^2,T^3$ are constrained as
\eqn\ExThreeT{\eqalign{&T^3 = {1\over2q^2}N_{1A}U^A =
{1\over2q^1}N_{2A}U^A = -{1\over2q^4}N_{4A}U^A,\cr
&q^2T^1+q^1T^2 = q^4T^4.}} 
Then, from $\partial_A W=0$, we have
\eqn\ExThreeU{N_{1A}T^1 + N_{2A}T^2 + N_{4A}T^4 = 0.}
Comparing the last set of equations to the second line of
Eq.~\ExThreeT, we see that solutions exist for
\eqn\ExThreeAnsatz{(N_{1A},N_{2A},N_{4A})= \l n_A
(q^2,q^1,-q^4).}
Here, $n_1$ and $n_3$ can be independently chosen to be $0$ or $1$, and
$\l$ is a proportionality constant.  (We take $n_0=n_2=0$ to remain in
case of parabolic twists as mentioned in the first paragraph.)  Using
Eq.~\ExThreeAnsatz, the first line of Eq.~\ExThreeT\ becomes
\eqn\ExThreeTnew{T^3 = \ha\l n_A U^A = \ha\l(n_1 U^1 + n_3 U^3).}

Let us write ${\bf q}=(-q^1,q^2,q^4)^T$.  Then two classes of
solutions with $\l=1$, and their associated contributions to the D6
Bianchi identities, are
\eqn\ExThreeClasses{\eqalign{%
{\bf q}^T &= 2(-1+z^2,2,2z),\quad z\in\IZ,\quad
-\half q^aN_{aA} = 8n_A,\cr
{\bf q}^T &= 2(-1+\half z^2,1,z),\quad z\in2\IZ,\quad -\half
q^aN_{aA} = 4n_A.}}
Additional examples with $\l=\half$ or $2$ are obtained from the
second class, by doubling ${\bf q}$~only or $N_{aA}$~only,
respectively.  In both cases, $-\half q^a N_{aA} = 8n_A$.  The number
of D6 branes wrapping 3-cycles of homology class $A^A$ is $8 + \half
q^a N_{aA}$.  This is either 0, 4, or 8 for each of the classes of
solutions that we have just described.

Note that from the $K3_{(3)}$ volume form, which is minus $(-T^1)T^2 +
T^2(-T^1) + \sum_I(T^{3I})^2$, we obtain a natural $SO(1,1+16)$
structure on these classes of solutions: the metric is of the form
\SOmetric, and $(-T^1,T^2,T^{3I})^T$ and $(-q^1,q^2,q^{3I})^T$
transform as vectors.  (This is the $SO(1,1+16)$ obtained from the
$SO(3,3+16)$ coset moduli space of $K3$ by restricting to K\"ahler
deformations only.)  Let us restrict to the $SO(1,1+1)$ acting on
$(-T^1,T^2,T^4)^T$ and ${\bf q}$. Then the classes above follow from
the $z=0$ class by $SO(1,1+1)$ rotation.  The matrix that implements
the rotation is
\eqn\exThreeSOmatrix{v = \pmatrix{1 &-\half z^2 & -z\cr
0 & 1 & 0\cr 0 & z & 0}.}
If we identify solutions that are related by action of the $K3$
modular group (i.e., the action of $SO(1,2;\IZ)\subset \G_{3,19}$),
then the first class collapses to just two solutions, corresponding to
$z=0$ and $1\hbox{ mod }2$. The second class collapses to one
solution.  The $z=0$ solutions are analogous to Example~1, with twists
inherited from $T^6$ only.  The only difference compared to Example~1
is that we permit $q^1$ and $q^2$ to be distinct, and we allow
$N_{aA}$ to be nonzero not only for $A=3$, but for $A=1$ as well.

\nonnumberedsubsec{Example 4:  Elliptic twists, but no flux}

Consider the choice of discrete data $q^a=0$, $-N_{11} = -N_{12} =
N_{21}=N_{22} = \pi$ and $N_{aA}=0$ otherwise.  From Eq.~\KthreeMNgen,
this choice corresponds to the $K3$ fibration data
\eqn\exFourdata{\eqalign{\g^3_{51} &=-\g^1_{53} = \pi,\cr
\b^{31}_5 &= -\b^{13}_5 = -\pi,\cr
h_{531} &= -h_{513} = -\pi.}}

Let us take a moment to describe the monodromies in this elliptic case
explicitly.  First, let $I$ and $S$ denote the $2\times2$ matrices
\eqn\IandS{I = \pmatrix{1 & \cr & 1}\quad\hbox{and}\quad
S = \pmatrix{ & -1\cr 1 & }.}
Then, the twist matrices \arbsoM\ corresponding to the choice
\exFourdata\ of $\b,\g,h$ are $M_6=0$ and
\eqn\exFourM{M_5 = \pi\pmatrix{S & -S\cr -S & S},}
where we have restricted to the nonzero $SO(2,2)$ block of $M_5$ with
indices $i=1,3$.

In the identification $SL(2,\IR)\times SL(2,\IR)\cong SO(2,2;\IR)$,
we have
\eqn\Sident{I\otimes S \cong \pmatrix{S & \cr & S},\quad
S\otimes I \cong \pmatrix{ & S\cr S & },\quad
S\otimes S \cong \pmatrix{ & -I\cr -I &  }.}
So, we can write
\eqn\MofS{M_5 \cong \pi(I\otimes S - S\otimes I).}
Using $S^2 = -1$, the matrix $\G_5(x^5) = \exp(-M_5x^5)$ appearing in
the definition of the twisted 2-forms \zetachi\ is 
\eqn\Gfive{\eqalign{\G_5(x^5) 
& \cong \exp(-\pi x^5 I\otimes S)\exp(\pi x^5 S\otimes I)\cr
& = \exp(\pi x^5 S)\otimes\exp(-\pi x^5 S)\cr
& = (\cos\pi x^5 + S\sin\pi x^5)\otimes (\cos\pi x^5 - S\sin\pi x^5).}}
The monodromy about the $x^5$ circle is $\G_5(1)$, while that about
each of the four fixed points of $\IP^1 = T^2(x^5,x^6)/\IZ_2$ is
$\G(\pm\half)$ (cf.~the discussions at the end of Sec.3.3 and in
Sec.~6.1).  From the last result, we have
\eqn\Gfiveof{\G_5(\pm\half) \cong -S\otimes S,\quad \G_5(1) = 1.}
Since $M_6$ vanishes, we have $\G_6(x^6) = 1$.

We now turn to moduli stabilization, as in the previous examples.  The
supersymmetry constraints on moduli are
\eqn\ExFourModconstr{T^1 = T^2\quad\hbox{and}\quad U^1 = U^2,}
from $\partial W/\partial U^A = 0$ and $\partial W/\partial T^a = 0$,
respectively.  The Minkowski condition $W=0$ is then satisfied
automatically.

Let us interpret these constraints in the orbifold limit of $\TZZ$,
setting all axionic moduli $B^a$ and $C^A_{(3)}$ equal to zero, for
simplicity.  The constraints become
\eqn\ExFoursmallr{r_1r_2 = r_3r_4\quad\hbox{and}\quad
r_1/r_2 = r_3/r_4,}
where the $r_i$ are the radii of $T^4_{(3)}$.  From
the discussion at the end of App.~A, the $K3$ moduli $G_{ij}$ are
related to $r_4$ and the metric $g_{ij}$ on $T^3(x^1,x^2,x^3)$ via
\eqn\ExFourG{G_{ij} = g_{ij}(r_4/\sqrt{g}) = g_{ij}/(r_1r_3).}
Thus,
\eqn\ExFourR{(R_1)^2 = r_1/r_3 = 1\quad\hbox{and}\quad 
(R_3)^2 = r_3/r_1 = 1.}
So, in the metric $G_{ij}$, the radii $R_1$ and $R_3$ are stabilized
to unity, by the nontrivial $SO(2,2)$ monodromy over $S^1(x^5)$.

This is identical to the stabilization of $T^2$ radii to unity in the
nongeometric ``T-fold''\foot{The D-brane spectrum in this 7D T-fold
background was recently studied in Ref.~\LawrenceMA.  For T-folds and
other duality twists of elliptic monodromy $(\G_n)^k =1$, Hellerman
and Walcher have given a complete 1-loop characterization of the full
string theory background as a generalization of the standard orbifold
construction.  As a special case, the particular T-fold just described
has been shown to yield a modular invariant partition function in type
II string theory and to preserve 16 supercharges
\HellermanTX .}  fibration of $T^2$ over $S^1$ with monodromy
\eqn\TfoldMonod{\r,\t\to -1/\r,-1/\t,}
where $\r$ and $\t$ are the K\"ahler and complex structure moduli of
the $T^2$.  In fact, this \hbox{T-fold} background and our twisted
$(K3\times T^2)/\IZ_2$ type IIA orientifold are dual---up to the
additional circles, orientifold and $\IZ_2\times\IZ_2$ orbifold
included in the latter \CveticHet.

\subsec{Supersymmetric Anti de Sitter vacua}

Now let us relax the condition that $W=0$ and consider generic
supersymmetric vacua.  The connection $K_{,\Phi^P}$ that appears in
the K\"ahler covariant derivatives is found to be
\eqn\KahlerConn{\eqalign{%
K_{,a} &= - {2\over\sqrt{8V_X}}\int\o_a\w J\w J,\cr
K_{,A} &= -2e^{K_2/2}\int\a_A\w\Im(C\hat\O)
= -2e^{K_2/2}\int\a_A\w\Im(e^{-\phi}\O).}}
If we define $\m = e^{K/2}W/\sqrt{8V_X}$, so that the cosmological
constant is $-|\mu|^2$, then the supersymmetry conditions become
\eqn\AdSsusyD{\eqalign{0 &= D_a W = i\int\o_a\w\bigl(
d\O_c + e^{J_c}\w\bar F_\RR + i\m e^{-2\phi}J\w J\bigr),\cr
0 &= D_A W = -i\int\a_A\w\bigl(
\bar H + dJ_c + 2i\m\Im(e^{-\phi}\O)\bigr).}}
Thus, 
\eqn\AdSsusy{\eqalign{%
& d\O_c + e^{J_c}\w\bar F_\RR + i\m e^{-2\phi}J\w J = 0,\cr
& \bar H + dJ_c + 2i\m\Im(e^{-\phi}\O) = 0,}}
with imaginary parts
\eqn\AdSsusyIm{\eqalign{%
&d(e^{-\phi}\Re\O) + J\w\tilde F_{(2)} + \Re(\m)e^{-2\phi}J\w J = 0,\cr
&dJ + 2\Re(\mu)e^{-\phi}\Im\O = 0,}}
and real parts
\eqn\AdSsusyRe{\eqalign{&\tilde F_{(4)} - \ha J\w J F_{(0)}
- \Im(\m)e^{-2\phi}J\w J = 0,\cr
&H - 2\Im(\m)e^{-\phi}\Im\O = 0.}}
We will not present examples of AdS vacua here, but note that the
general equations \AdSsusy\ were studied in Ref.~\GranaSN.  Examples
in closely related contexts appear in Refs.~\refs{\BehrndtMJ,\LustIG}.


\newsec{Geometric twists for Calabi-Yau manifolds of Voisin-Borcea type}

The class of Voisin-Borcea Calabi-Yau manifolds of the form $(K3\times
T^2)/\IZ_2$ \BorceaTQ\ contains many more manifolds that $\TZZ$.  This
class is discussed in more detail in App.~C.  For our purposes here,
we note that each Voisin-Borcea Calabi-Yau manifold is defined, in
part, by $\IZ_2$ involution of the second cohomology lattice
\eqn\Htwolattice{H^2(K3,\IZ) = (- E_8\times E_8)\times U_{1,1}^3.}
Here, $U_{1,1}$ is a two dimensional lattice of signature $(1,1)$.  For
the $\TZZ$ case above, we have
\eqn\Hplusminus{\eqalign{%
H^2_{+} &= (- E_8\times E_8)\times U_{1,1}
\qquad\hbox{(from $\chi_I$; $\chi_3$, $\chi^3$),}\cr
H^2_{-} &= (U_{1,1})^2\qquad\quad
\qquad\hbox{(from $\chi_1$, $\chi^1$; $\chi_2$, $\chi^2$).}}}
However, the twists described above easily extend to other
involutions, which lead to different $H^2_{+}$ and $H^2_{-}$.
Following standard notation, we let $r$ denote the rank of the even
part of the cohomology lattice $H^2_{+}$.

Let us assume for simplicity that the $U_{1,1}^3$ factor in $H^2$
still decomposes into $U_{1,1}$ in $H^2_{+}$ and $U_{1,1}^2$ in
$H^2_{-}$.  Then, the parity of the $\chi_i$ and $\chi^i$ is
unchanged, but the $\chi_I$ of the $E_8\times E_8$ factor in general
decompose into $r-2$ elements $\chi_{I^+}$ of $H^2_{+}$ and $18-r$
elements $\chi_{I^-}$ of $H^2_{-}$.  (For $\TZZ$, we have $r=18$ and
all $\chi_I$ are in $H^2_{+}$.)  So, the result~\ModuliTwists\
generalizes to
\eqn\ModuliTwistsPlus{\eqalign{%
\hbox{Moduli:}\quad & \d A^{I^+3},\cr 
\hbox{Twists:}\quad & A^{I^+i} = m^{I^+i}_n x^n,\quad i=1,2,\quad n=5,6,
\quad m^{I^+i}_n\in4\IZ,}}
and
\eqn\ModuliTwistsMinus{\eqalign{%
\hbox{Moduli:}\quad & \d A^{I^-1},\ \d A^{I^-2},\cr 
\hbox{Twists:}\quad & A^{I^-3} = m^{I^-3}_n x^n,\quad n=5,6,
\quad m^{I^-i}_n\in4\IZ.\hphantom{\quad i=1,2,}}}

Here we have assumed the parabolic case, in the terminology of
Sec.~3.3.  The twist data $m^{I^+i}_n$ and $m^{I^-3}_n$ again
partially characterizes the fibration of $K3$ over $T^2$.
Eqs.~\ModuliTwistsPlus\ and \ModuliTwistsMinus\ indicate which
components are compatible with the $\IZ_2$ involution.  The subset of
$\g^i_{nj}$ and $\b^{ij}_n$ preserved by the involution is the same as
that for $\TZZ$: we require one of $i,j$ to be $3$ and the other to be
$1$ or $2$.  To ensure that the $\IZ_2$ quotient of the $K3$ fibration
over $T^2$ remains well defined as a fibration over $\IP^1=T^2/\IZ_2$,
we again require that $\g^i_{nj}$, $\b^{ij}_n$, $\half m^{I^+i}_n$ and
$\half m^{I^-3}_n$ be even integers.

For the basis of differential forms, we choose a very similar basis to
that of $\TZZ$.  For the 2-forms $\o_a$, the two differences are: (1)
the $K3$ blow-up modes $\o_{3I^+}$ are now labeled by $I^+$ only, and
(2) instead of $\o_{1I}$ and $\o_{2I}$, we now have an
involution-dependent set of 2-forms $\o_{I'}$, that result from
blowing up the singularities of $(K3\times T^2)/\IZ_2$.  For the
3-forms $\a_A$ and $\b^A$, the analogous differences are: (1) we now
have 3-forms from the $K3$ complex structure deformations generated by
by the $\z_{I^-}$, for which choose the symplectic basis
\eqn\VBthreeforms{\a_{I^-} = -dx^5\w\z_{I-},\quad
\b^{J^-} = dx^6\w\z_{I^-}\d^{I^-J^-},}
and (2) in addition, we have an involution-dependent set of 3-forms
that generate complex structure deformations of the singularities of
$(K3\times T^2)/\IZ_2$.

Without any assumption on whether the twists are parabolic, elliptic
or otherwise, the result \KthreeMNgen\ generalizes to
\eqn\VBM{M_a{}^A = \pmatrix{%
\left.\matrix{\g^2_{53} & -h_{631} & \g^1_{63} & -h_{532}
\vphantom{n^{K^-}_{53}}\cr
\noalign{\vskip2pt}
\b^{32}_5 & -\g^3_{61} & \b^{31}_6 & -\g^3_{52}
\vphantom{m^{K^-3}_5}\cr
\noalign{\vskip2pt}
0 & 0 & 0 & 0\cr
\noalign{\vskip6pt\hrule\vskip6pt}
0 & 0 & 0 & 0\cr
m^{J^+2}_5 & n^{J^+}_{61} & m^{J^+1}_6 & n^{J^+}_{52}}\ \right|\,
\matrix{0 & n^{K^-}_{53}\cr 
\noalign{\vskip2pt}
0 & m^{K^-3}_5\cr
\noalign{\vskip2pt}
0 & 0\cr 
\noalign{\vskip6pt\hrule\vskip6pt}
0 & 0\cr 0 & f^{J^+K^-}_5}}}
and
\eqn\VBN{N_{aA} = \pmatrix{%
\left.\matrix{h^{632} & \g^1_{53} & -h_{531} & -\g^2_{63}
\vphantom{-n^{K^-}_{63}}\cr
\noalign{\vskip2pt}
\g^3_{62} & \b^{31}_5 & \g^3_{51} & -\b^{32}_6
\vphantom{m^{K^-3}_6}\cr
\noalign{\vskip2pt}
0 & 0 & 0 & 0\cr
\noalign{\vskip6pt\hrule\vskip6pt}
0 & 0 & 0 & 0\cr
-n^{J^+}_{62} & m^{J^+1}_5 & n^{J_+}_{51} & -m^{J^+2}_6}\ \right|\,
\matrix{\ 0 & -n^{K^-}_{63}\cr 
\noalign{\vskip2pt}
\ 0 & m^{K^-3}_6\cr
\noalign{\vskip2pt}
\ 0 & 0\cr 
\noalign{\vskip6pt\hrule\vskip6pt}
\ 0 & 0 \cr \ 0 & f^{J^+K^-}_6}}.}
Here, the vanishing rows correspond to $\o_3$ and the 2-forms of type
(2) above.  The vanishing column corresponds to the 3-forms of type
(2) above.


\newsec{Further generalizations}

\subsec{Geometric twists of other K3 fibered Calabi-Yau manifolds}

Having discussed the geometric twists of $X=(K3\times T^2)/\IZ_2$ from
the point of view of the $K3\times T^2$ covering space, let us now
describe the twists more intrinsically from the point of view of the
resulting Calabi-Yau quotient $X$.  As we will see, this leads to a
natural generalization beyond the Voisin-Borcea class.  For generic
points on the $T^2$ in the covering space, the $K3$ surface at a point
$(x^5,x^6)$ on $T^2$ is identified with another $K3$ surface at the
point $(-x^5,-x^6)$ via the $\IZ_2$ involution.  Thus, the quotient
$X$ is a fibration over $T^2/\IZ_2 = \IP^1$, with generic fiber $K3$
and with singular fibers at the four fixed points on the base.  This
statement remains true if we instead begin with a nontrivial fibration
of $K3$ over $T^2$ as in the twisted geometry.

The twists of Eqs.~\ModuliTwistsPlus\ and \ModuliTwistsMinus\ can be
equivalently expressed as the monodromies of the $K3$ moduli
$A^{I^+}_i$ and $A^{I^-}_i$ about the $x^5$ and $x^6$ circles in $T^2$:
\eqn\MonodromyTtwo{\eqalign{%
&A^{I^+i} \to A^{I^+i}+m^{I^+i}_5\quad\hbox{and}\quad
A^{I^-3} \to A^{I^-3}+m^{I^-3}_5\quad\hbox{for}\quad
x^5\to x^5+1,\cr
&A^{I^+i} \to A^{I^+i}+m^{I^+i}_6\quad\hbox{and}\quad
A^{I^-3} \to A^{I^-3}+m^{I^-3}_6\quad\hbox{for}\quad
x^6\to x^6+1.}}
After performing the $\IZ_2$ identification, the $x^5$ and $x^6$
circles cease to exists as homology cycles in the $\IP^1$, but
monodromies in Eq.~\MonodromyTtwo\ survive as monodromies about the
four $\IZ_2$ fixed points.

\medskip\centerline{\figscale{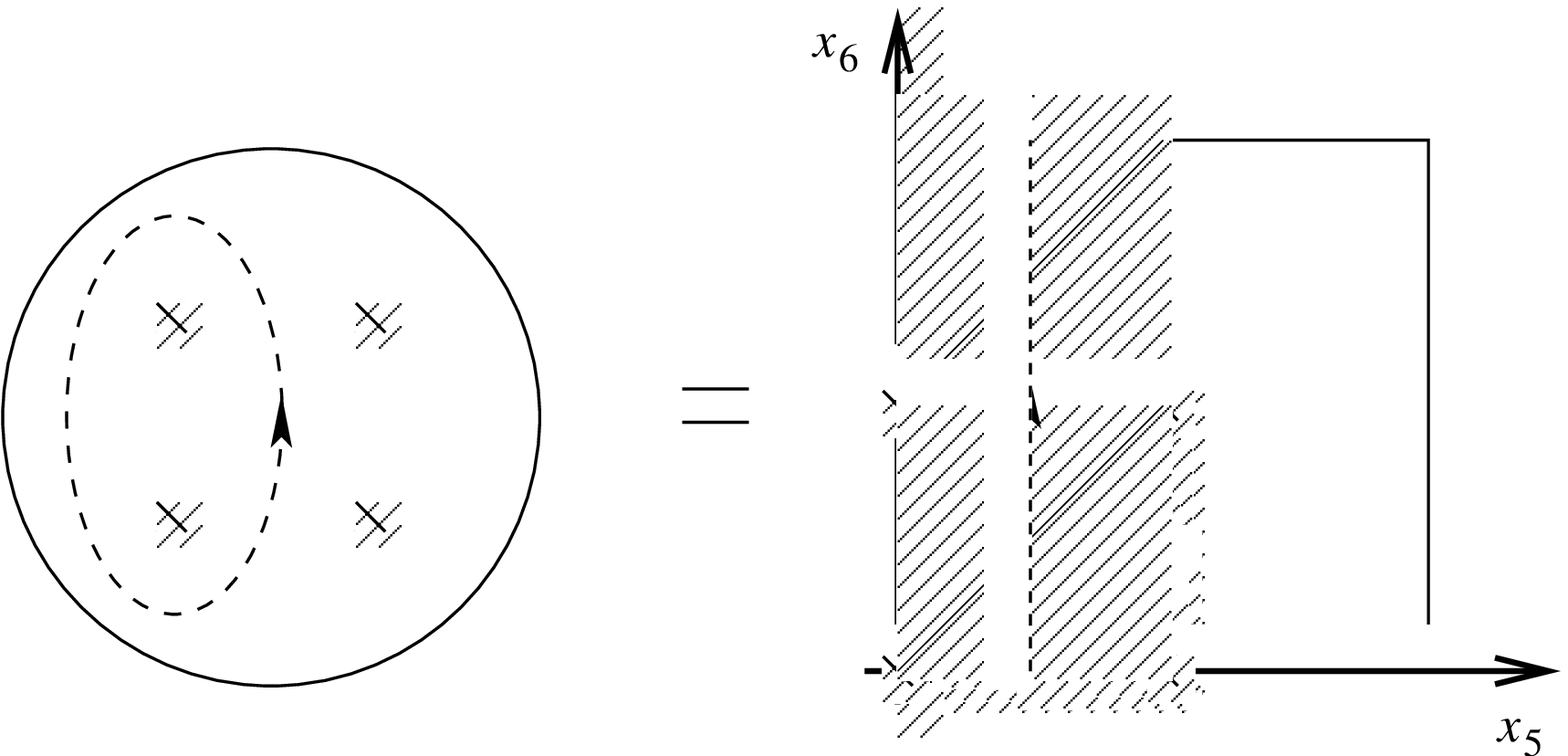}{4.0truein}}\nobreak
\caption{1}{In the identification $\IP^1$ (left) $= T^2/\IZ_2$ (right),
the monodromies of the $K3$ moduli about homologically nontrivial
circles in $T^2$ become monodromies about the $\IZ_2$ fixed points in
$\IP^1$.}\smallskip

Let $p_1$, $p_2$, $p_3$ and $p_4$ denote the four fixed points with
$T^2$ coordinates $(0,0)$, $(\half,0)$, $(\half,\half)$ and
$(0,\half)$, respectively (i.e., counterclockwise starting in the
lower left in Fig.~1).  Then, the monodromies about these fixed points
are
\eqn\MonodromyPone{\eqalign{%
M_1:\quad & 
A^{I^+i} \to A^{I^+i}+\half m^{I^+i}_5+\half m^{I^+i}_6
\quad\hbox{and}\quad
A^{I^-3} \to A^{I^-3}+\half m^{I^-3}_5+\half m^{I^-3}_6,\cr
M_2:\quad & 
A^{I^+i} \to A^{I^+i}+\half m^{I^+i}_5-\half m^{I^+i}_6
\quad\hbox{and}\quad
A^{I^-3} \to A^{I^-3}+\half m^{I^-3}_5-\half m^{I^-3}_6,\cr
M_3:\quad & 
A^{I^+i} \to A^{I^+i}-\half m^{I^+i}_5-\half m^{I^+i}_6
\quad\hbox{and}\quad
A^{I^-3} \to A^{I^-3}-\half m^{I^-3}_5-\half m^{I^-3}_6,\cr
M_4:\quad & 
A^{I^+i} \to A^{I^+i}-\half m^{I^+i}_5+\half m^{I^+i}_6
\quad\hbox{and}\quad
A^{I^-3} \to A^{I^-3}-\half m^{I^-3}_5+\half m^{I^-3}_6.}}
As a check, note that $M_1 M_4$ gives the monodromy
$$M_1M_4:\quad A^{I^+i} \to A^{I^+i}+m^{I^+i}_6
\quad\hbox{and}\quad A^{I^-3} \to A^{I^-3}+m^{I^-3}_6,$$
which is indeed the correct monodromy about the $x^6$ circle of
Fig.~6.  Likewise $(M_2 M_3)^{-1}$ gives the monodromy about the $x^6$
circle, while $M_1M_2$ and $(M_3M_4)^{-1}$ give the correct monodromy
about the $x^5$ circle.  More generally, given monodromies
$\G_5(1)=\bigl(\G_5(\half)\bigr)^2$ and
$\G_6(1)=\bigl(\G_5(\half)\bigr)^2$ about the $x^5$ and $x^6$ circles
of $T^2$, the monodromies about the four fixed points are
\eqn\MondoromyPoneGen{\eqalign{&\G_{p_1} = \G_5(\half)\G_6(\half),
\quad\G_{p_2} = \G_5(\half)\G_6(-\half),\cr
&\G_{p_3} = \G_5(-\half)\G_6(-\half),\quad
\G_{p_4} = \G_5(-\half)\G_6(\half).}}

There is a natural generalization to other $K3$ fibered Calabi-Yau
manifolds.  For any such fibration, there is a set of points $p_i$ on
the base $\IP^1$ over which the $K3$ fiber degenerates, and an
associated set of $\G_{3,19}$ monodromies $M_i$ that give the
automorphisms of the homology lattice of the K3 fibers that result
from circling these points.  We can alter these monodromies.  Provided
that the total monodromy about all of the $p_i$ is trivial, we obtain
another well defined manifold, which in general is non Calabi-Yau.

A very similar idea was discussed by in Ref.~\TomasielloBP, in the
case of mirror of the quintic hypersurface in $\IP^4$, viewed as a
$T^3$ fibration over $S^3$.  Here, the twists instead represent
modified monodromies about singular loci on the $S^3$.  This idea
extends in principle to other Calabi-Yau manifolds, provided that the
Strominger-Yau-Zaslow $T^3$ fibration is known.  However, the
degenerate locus on the base in this case consists of a one
dimensional web in $S^3$ as opposed to a collection of points in
$\IP^1$.  (For the quintic, this web is known and has been described
by M.~Gross \GrossHC, as summarized in \TomasielloBP.)  Therefore, in
practice, explicit realizations seem easier in the $K3$ fibered
context described here.  Of course, for Calabi-Yau manifolds that are
both $K3$ and $T^3$ fibered, the two constructions should agree.  Note,
however, that $\TZZ$ does not fall into this category since the mirror
is not purely geometric, but is instead an orbifold with discrete
torsion.

\subsec{Nongeometric twists}

Let us return to the $\IZ_2$ covering space description of the twists
for the Voisin-Borcea class.  The geometric twists are defined by a
nontrivial fibration of $K3$ over $T^2$.  As discussed in Sec.~3.3,
compactification of string theory or supergravity on the twisted
geometry can be viewed as a Scherk-Schwarz compactification---a two
step compactification, in which the moduli from the first step of the
compactification (on $K3$) are given nontrivial dependence on the
coordinates of the second second step of the compactification (on
$T^2$).\foot{Here, we use the term Scherk-Schwarz {\it
compactification} as opposed to Scherk-Schwarz {\it reduction\/} to
distinguish the full twisted compactification from its truncation to a
particular subsector.  See Ref.~\HullHK\ and the Introduction of
Ref.~\HullTP\ for a discussion of this point.  In their terminology, a
two step compactification, with a clear fiber theory and base, would be
termed a duality twist.  The twisted tori discussed at the beginning
of Sec.~3.2 are not necessarily of this type, in that they include,
for example, the $SU(2)$ group manifold $S^3$ for $\g^i_{jk} =
\e^i{}_{jk}$.  Although $S^3$ can be viewed as a Hopf fibration of $S^1$
over $S^2$, when viewed as a twisted $T^3$ it has no clear fiber and
base: the $\g^i_{jk}$ parametrize twists of all three $S^1$ factors
over all three complementary $T^2$ factors.}  Compactification of type
IIA string theory or supergravity on $K3$ includes more than just
metric moduli.  The massless spectrum consists of the $\CN=(1,1)$ 6D
gravity multiplet and 20 vector multiplets.  The vector multiplets
each contain 4 scalars, so there are a total of 80 scalars in matter
multiplets.  Of these, 58 are accounted for by the $K3$ metric moduli
space \ModKthree\ and 22 by the moduli space $H^2(K3,U(1))\sim
H^2(K3,\IR)/H^2(K3,\IZ)$ of the NS $B$-field.  The one remaining
scalar is the dilaton, in the gravity multiplet.

After accounting for all discrete identifications, the complete moduli
space for type IIA compactification on $K3$ is\foot{In fact, there is
a beautiful interpretation of this moduli space as the space of
positive signature 4-plane in the total cohomology space
$H^*(K3,\IR)=\IR^{4,20}$.  See, for example, 
Refs.~\refs{\WendlandRY,\AspinwallMN}.} 
\eqn\IIAModKthree{\CM_{\rm IIA} = \IR_{>0}\ \times\ {} 
\bigl(SO(4)\times SO(20)\bigr)\backslash SO(4,20)/\G_{4,20}.}
In the second step of the Scherk-Schwarz compactification,
compactification on $T^2$, we need not restrict ourselves to the
geometric duality group of the fiber theory, $\G_{3,19}$, but can
instead allow are arbitrary commuting monodromies
$\G_5(1),\G_6(1)\in\G_{4,20}$ under $x^5\to x^5+1$ and $x^6\to x^6+1$.
Since these monodromies in general mix the metric and $B$-field moduli,
the compactifications are in general nongeometric \HellermanAX.  This
construction is similar to the T-fold construction of Ref.~\HullIN.
In fact there should be a similar, ``partially doubled geometry''
description of these compactifications analogous to the doubled torus
of \HullIN.  To linearize the action of the duality group $\G_{4,20}$,
only the part inherited from $T^4$, contributing duality group
$\G_{4,4}$, actually needs to be doubled.

A still more general nongeometric construction, would be to allow
$\G_{4,20}$ monodromies not only over the torus $T^2(x^5,x^6)$, but
simultaneously over the T-dual torus $T^2(\tilde x^5,\tilde x^6)$.
Constructions of this type have been discussed in Ref.~\DabholkarVE.
Again, both here and in the previous paragraph, there is a natural
generalization to compactifications on arbitrary $K3$ fibered
Calabi-Yau manifolds.

\subsec{Comments on effective field theory}

A critical property for all of the twisted compactifications discussed
here is that there is a natural split into fiber and base.  For the
$\CN=2$ pre orientifold theory, the $K3$ fibration over $T^2$ gives
rise to precisely the right twist data to parametrize couplings of
vector multiplets to hypermultiplets in an $\CN=2$ gauged supergravity
theory.  In the \hbox{$(K3\times T^2)/\IZ_2$} orbifold untwisted
sector, the vector multiplet moduli space is classically an
$SL(2)\times SO(2,r)$ coset and the hypermultiplet moduli space is a
\hbox{$SO(4,22-r)$} coset.  (Here, $r$ is the Voisin-Borcea parameter
introduced in Sec.~5 and App.~B, with $r=18$ for $\TZZ$.)  The
couplings are the $\bigl(SL(2)\times SO(2,r),SO(4,22-r)\bigr)$
bifundamentals whose moduli are projected out by the orbifold $\IZ_2$.
For $T^2$ fibrations over $K3$ there is likewise a natural split.

From the point of view of the parent theory on $T^2\times K3$, there
is no such natural split in the $\CN=4$ low energy effective theory,
despite the fact that there is one in the target space geometry.  The
moduli space is an $SL(2)\times SO(6,6+16)$ coset, as is clear from
the dual heterotic description on $T^6$.  Consequently, the gauged
$\CN=4$ supergravity data for the $K3$ fibration over $T^2$
represents, from the point of view of the effective field theory, an
arbitrary choice of decomposition:
\eqn\NfourSplit{SL(2)\times SO(6,6+16)\quad\to\quad
\bigl(SL(2)\times SO(2,r)\bigr)\ \times\ SO(4,22-r).}

Finally, note that the construction by Tomasiello \TomasielloBP\
employs a different fiber/base split of $T^3$ over $S^3$.  Thus,
although there is an overlapping class of twisted geometries that are
simultaneously $K3$ fibrations over $\IP^1$ and $T^3$ fibrations over
$S^3$, both constructions should contain twists that the other does
not, and therefore give somewhat different possible couplings in the
low energy effective field theory.


\newsec{Conclusions}

We have described how discrete geometric twists can be included in the
set of defining data for string compactifications based on Calabi-Yau
manifolds of Voisin-Borcea type $(K3\times T^2)/\IZ_2$.  The twist
data define a nontrivial fibration of $K3$ over $T^2$ compatible with
the $\IZ_2$ involution, and a corresponding discrete deformation of
the closure and exactness relations of the Calabi-Yau cohomology ring.
The data can be parametrized either in terms of monodromies $\G_n$ in
the automorphism group $\G_{3,19}$ of the $K3$ cohomology lattice
(under $x^n\to x^n+1$ on the $T^2$), or their logarithms, which appear
in the modified closure relations.  The quantization conditions on the
latter depend on the conjugacy class of $\G_n$, as we have
illustrated through two concrete cases.  We have termed these cases
parabolic and elliptic by analogy to the conjugacy classes of
$SL(2,\IR)$.

For the particular Voicin-Borcea manifold $\TZZ$, which features
prominently in model building, we have studied the type IIA
orientifold of the twisted background in detail.  Supersymmetric
Minkowski vacua are of the holomorphic monopole
form~\refs{\KasteXS,\GranaBG} and lift to M~theory compactified on
manifolds of $G_2$ holonomy.  We have presented four examples of such
vacua, three of parabolic type with nonzero Ramond-Ramond flux and one
of elliptic type with no flux.  A feature that we observed is that
even when the discrete data involves the orbifold untwisted sector
only, the blow-up moduli can nevertheless be stabilized due to the
interplay between orbifold untwisted and twisted sectors in the
intersection form of the Calabi-Yau.  The first three examples share
the property that all of the blow-up K\"ahler moduli of $(K3\times
T^2)/\IZ_2$ are classically stabilized to zero.  By analogy to
conformal field theory results in the context of the heterotic string
on $T^6/\IZ_3$ \CveticAC, we suspect that this result persists in the
presence of worldsheet instanton corrections.  However, this clearly
needs to be explored further.  We are currently investigating the
requisite CFT expansion about the orbifold limit, for small blow-up
K\"ahler moduli.

There are several other possible directions for future work.  First,
the geometric twists constructed here include only one third of the
blow-up modes of $\TZZ$: those that resolve $T^4_{(3)}/\IZ_2$ to
$K3_{(3)}$, to give $(K3\times T^2)/\IZ_2$.  Our construction can
never twist the exceptional cohomology of $(K3\times T^2)/\IZ_2$.  As
discussed in Sec.~6, the monodromies introduced here are monodromies
{\it around\/} the $\IZ_2$ fixed points on the base $\IP^1 =
T^2/\IZ_2$, while the exceptional cohomology is associated with
singular fibers localized {\it at\/} the fixed points on the base.  It
would be interesting to find a different way to twist the geometry
that can include the $(K3\times T^2)/\IZ_2$ blow-up modes and
exceptional complex structure deformations.

While the main goal of our work was to construct, for the
Voisin-Borcea class, the geometric twists described in Sec.~2, we were
soon confronted with the larger challenge of understanding the class
of type IIA Calabi-Yau orientifolds with Ramond-Ramond flux and
geometric twists.  There is currently a gap in the literature in terms
of even qualitatively understanding vacua of this type.  Nevertheless,
this class is the natural geometric analog in type IIA of Calabi-Yau
orientifolds in type IIB with Neveu-Schwarz and Ramond-Ramond fluxes.
These IIA and IIB classes becomes mirror to one another only after
nongeometric twists are included as well, however, the simplest
subclasses that one could hope to understand are the geometric
subclasses for either IIA and IIB.  Much work has been done for IIB
with fluxes but very little for IIA with flux and twists (see for
example the reviews \refs{\DouglasES,\BlumenhagenCI}).  This is a
clear avenue for future work.

Finally, a great deal of our understanding of the space of string
theory vacua has come from duality, with type IIA/heterotic duality
featuring prominently in our understanding of type IIA vacua based on
$K3$ fibrations.  Since the central aim of our work was to understand
the discrete geometric data that can be incorporated into vacua based
on K3 fibrations, it is natural to ask what the dual heterotic
description of this data is.  As discussed in Sec.~3.3, the geometric
twists of $(K3\times T^2)/\IZ_2$ analyzed here represent
Scherk-Schwarz twists of $K3$ moduli upon further compactification on
the $T^2$.  For the twists surviving the $\IZ_2$ projection, the
corresponding continuous moduli are projected out.  In the heterotic
dual, one analog of this is heterotic flux for which the orbifold
projects out the corresponding zero modes of the gauge fields.  The
complete duality map and dual heterotic description is currently
under investigation \CveticHet.


\bigskip\centerline{\bf Acknowledgements}\nobreak\medskip\nobreak

It is a pleasure to thank V.~Balasubramanian, A.~Grassi, T.~Grimm,
A.~Neitzke and A.~Tomasiello for helpful discussions, as well as
T.~Weigand and K.~Wendland for useful references.  In addition, MC and
MBS thank the Aspen Center for Physics and the Kavli Institute for
Theoretical Physics for hospitality and a stimulating environment
during the course of this work.  This work was supported in part by
the DOE under contract DE-FG02-95ER40893 and the National Science
Foundation under Grant No.~PHY99-07949.


\appendix{A}{HyperK\"ahler structure on $T^4$}

A choice of hyperK\"ahler structure on $T^4$ is analogous to a choice
of complex structure on $T^2$.  Let us first review the latter in a
way that makes the generalization natural, and then go on to discuss
hyperK\"ahler structure on $T^4$.  This review is taken more or less
directly from Ref.~\SchulzKK, currently in preparation by one of the
authors.

On $T^2$, we can express the metric as
\eqn\Ttwoframemetric{ds^2_{T^2} = e^1\otimes e^1 + e^2\otimes e^2,}
where $e^m = e^m{}_n dx^n$ in terms of the vielbein $e^m{}_n$.  The
complex structure is defined by a tensor $J_i{}^j$ which we view as a
map\foot{The usual convention in the math literature is the transpose
of this: $J$ has index structure $J^m{}_n$, so that $J$ acts from the
left on the tangent space, and $J^T$ acts from the left on the
cotangent space.  However, if we require that (i) $J$ with holomorphic
(antiholomorphic) indices be $+i$ ($-i$), as is customary in both the
math and physics conventions, and (ii) the K\"ahler form be obtained
by lowering one index of the tensor $J$, with no sign change, then we
are uniquely led to the conventions used in this paper.}  $J\colon\
T^*\to T^*$, such that
\eqn\Jmap{J\colon\quad e^2\to e^1,\quad e^1\to -e^2.}
Lowering the upper index of $J_m{}^n$ gives the K\"ahler form on
$T^2$.  By $SL(2,\IZ)$ change of lattice basis for the lattice $\L$
that enters into $T^2 = \IR^2/\L$, we can always write
\eqn\Ttwovielbein{\eqalign{e^1 &= R^1(dx^1+a^1{}_2 dx^2),\cr
e^2 &= R^2 dx^2,\quad\hbox{where} \quad x^n\cong x^n+1.}}
The holomorphic 1-form is
\eqn\HoloOneform{e^z = e^1+ie^2 = R^1(dx^1+\t_1 dx^2),}
where $\t_1 = a^1{}_2 + i{R^2\over R^1}$ is the complex structure
modulus.

Likewise, we can express the metric on $T^4$ as
\eqn\Tfourframemetric{ds^2_{T^4} = e^1\otimes e^1 + e^2\otimes e^2
+ e^3\otimes e^3 + e^4\otimes e^4,}
where, again, $e^m = e^m{}_n dx^m$ in terms of the vielbein $e^m{}_n$
The hyperK\"ahler structure is defined by a triple of tensors
$J_{(i)m}{}^n$, $i=1,2,3$, which we view as maps $J_{(i)}\colon\
T^*\to T^*$, such that
\eqn\JHKmap{\eqalign{J_{(1)}\colon\quad &
e^4\to e^1,\quad e^3\to e^2,\quad e^1\to -e^4,\quad e^2\to -e^3,\cr
J_{(2)}\colon\quad &
e^4\to e^2,\quad e^1\to e^3,\quad e^2\to -e^4,\quad e^3\to -e^1,\cr
J_{(3)}\colon\quad &
e^4\to e^3,\quad e^2\to e^1,\quad e^3\to -e^4,\quad e^1\to -e^2.}}
The $J_{(i)m}{}^n$ satisfy 
$$J_{(1)} J_{(2)} = - J_{(2)} J_{(1)} = -J_{(3)},\qquad (J_{(1)})^2 = -1,$$
plus cyclic permutations.  Lowering the upper index on $J_{(i)m}{}^n$
gives a triple of K\"ahler forms $J_{(i)mn}$.  The quaternionic 1-form
is
\eqn\QuatOneform{e^q = e^4 - {\bf i} e^1 - {\bf j} e^2-{\bf k} e^3,}
where the quaternions ${\bf i},{\bf j},{\bf k}$ satisfy the same
algebra as $-J_{(i)}$:\foot{Here, $-J_{(i)}$ rather than $J_{(i)}$
satisfies the quaternion algebra for the reason discussed in the
previous footnote.  Note that the tangent space map $J^T_{(i)}$
satisfies the quaternion algebra with no minus sign.}
\eqn\QuatAlg{{\bf i}{\bf j} = {\bf k},
\quad {\bf j}{\bf k} = {\bf i},
\quad{\bf k}{\bf i} = {\bf j},\quad\hbox{and}\quad 
{\bf i}^2 = {\bf j}^2 = {\bf k}^2 = -1.}

A choice of complex structure on $T^4$ is then a choice of $i$ on the
${\bf i},{\bf j},{\bf k}$ unit sphere.  By a $SL(4,\IZ)$ change of
lattice basis for the $T^4$, we can write, in addition to
Eq.~\Ttwovielbein,
\eqn\Tfourvielbein{\eqalign{%
e^3 &= R^3(dx^3 + a^3{}_1 dx^1 + a^3{}_2 dx^2),\cr
e^4 &= R^4(dx^4 + a^4{}_1 dx^1 + a^4{}_2 dx^2 + a^4{}_3 dx^3),
\quad\hbox{where} \quad x^n\cong x^n+1.}}
So, for example, if we choose complex structure $i={\bf k}$, then the
complex pairing that follows from $J = J_{(3)}$ is
\eqn\Tfourcpx{\eqalign{%
e^{z^1} &= R_1 e^1 + i R_2 e^2 = R^1 (dx^1 + \t_1 dx^2),\cr
e^{z^2} &= R_4 e^4 - i R_3 e^3 = R^4 (dx^4 +\t_2{}^{-1} dx^3 + \ldots),}}
where $\t_2{}^{-1} = a^4{}_3 - i{R_3\over R_4}$ and the ``\dots'' is a
1-form on $T^2(x^1,x^2)$, which can be interpreted as the connection
for a trivial fibration of $T^2(x^3,x^4)$ over $T^2(x^1,x^2)$.  The
holomorphic $(2,0)$ form in this case is
\eqn\twozeroform{\O_{(2,0)} = e^{z^1}\w e^{z^2} = J_{(1)} + iJ_{(2)}.}

If we write the metric on $T^4$ as
\eqn\Tfourmetric{ds^2_{T^4} = R_4{}^2 (dx^4 + a^4_i dx^i)^2
+ g_{ij} dx^i dx^j,\quad i=1,2,3,}
then the choice of hyperK\"ahler structure is the choice of $a^4{}_i$
together with the dimensionless metric $G_{ij} =
(R_4/\sqrt{\mathstrut g})g_{ij}$.  Let us define $B^{ij} = -a^4{}_k\e^{kij}$,
where $\e^{123} = 1$.  Then, this choice parametrizes the
$\bigl(SO(3)\times SO(3)\bigr)\backslash SO(3,3)/\G_{3,3}$ truncation
of the coset \SOvielbein, with vielbein
\eqn\SOvbTrunc{V = \pmatrix{E & -EB\cr 0 & E^{-1T}},}
where $E^\l{}_i$ is the vielbein for the metric $G_{ij}$.  The coset
can also be interpreted as the choice of positive signature 3-plane
spanned by $J_{(1)},J_{(2)},J_{(3)}$ in $H^2(T^4,\IR) = \IR^{3,3}$.


\appendix{B}{The homology lattices of $T^4/\IZ_2$ and $\TZZ$}

For completeness, we review the integer homology lattices of
$T^4/\IZ_2$ and $\TZZ$.  This review is based primarily on
Refs.~\DenefMM\ and \WendlandRY.

\nonnumberedsubsec{The lattice of $T^4/\IZ_2$}

Let us view $T^4$ as $T^2_{(1)}(x^1,x^2) \times T^2_{(2)}(x^3,x^4)$
with complex pairing $dz_1 = dx^1+\t_1 dx^2$ and $dz_2 = dx^3+\t_2
dx^4$.  Now consider $T^4/\IZ_2$.  There are $2^4=16$ points of local
geometry $\IC^2/\IZ_2$ (16 $A_1$ singularities), located at the fixed
points where each of the four coordinates is equal to $0$ or $1/2$.
There are also $4+4=8$ fixed lines $\IP^1$ with a simple description
in this complex structure: let $D_{1s}$, $s=1,2,3,4$ label the
divisors $\IP^1 = T^2_{(2)}/\IZ_2$ located at each of the four fixed
points in $(x^1,x^2)$ and $D_{2t}$ denote the divisors $\IP^1 =
T^2_{(1)}/\IZ_2$ located at the four fixed point in $(x^3,x^4)$.  The
intersections of these $\IP^1$s in the singular geometry is
illustrated schematically in Fig.~2~(a).

\bigskip\centerline{\figscale{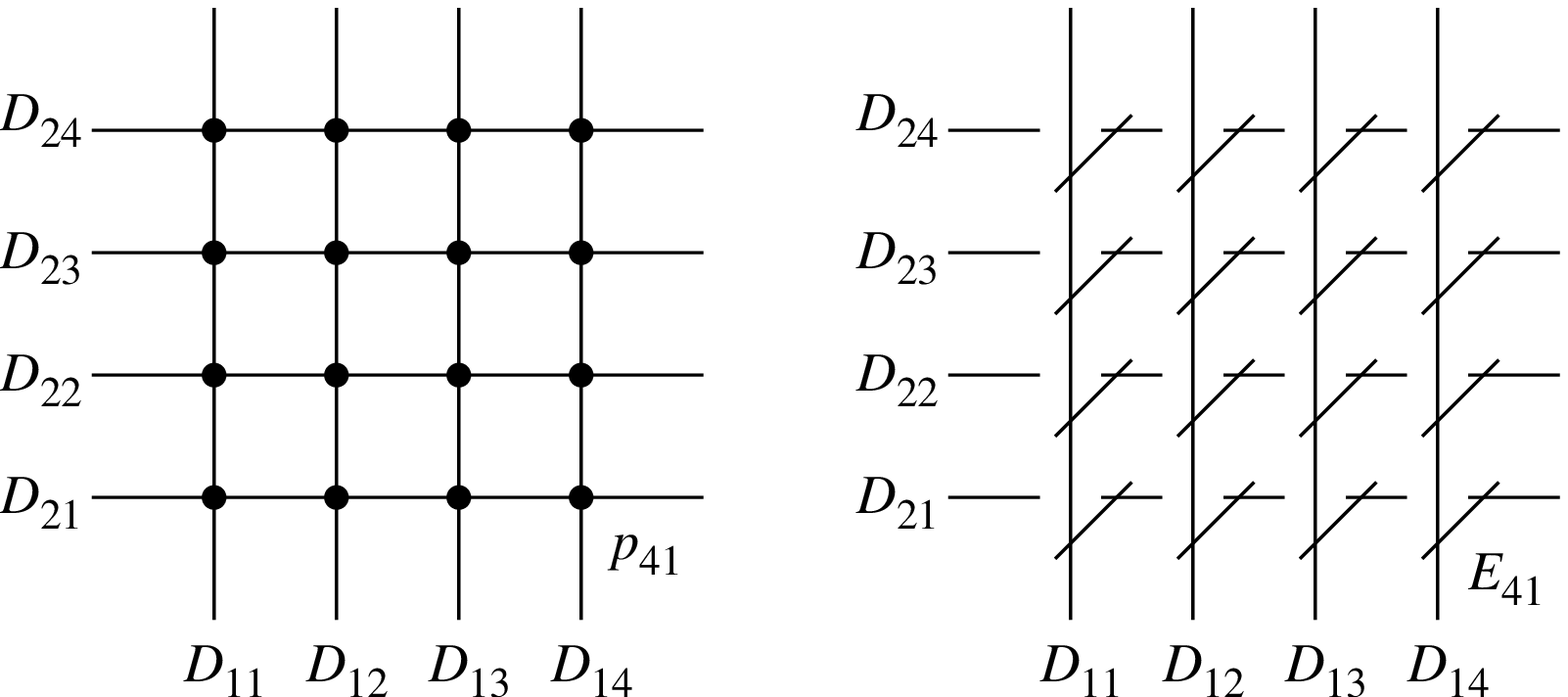}{4.0truein}}\nobreak
\caption{2}{(a) In the singular $T^4/\IZ_2$ (left), each of the sixteen
$A_1$ singularities is the ``half point'' of intersection, $p_{st}$,
of two fixed $\IP^1$s, $D_{1s}$ and $D_{2t}$.  (b) In the resolved
$K3$ (right), each $p_{st}$ is blown up to an exceptional divisor
$E_{st}$.  After resolution, $D_{1s}$ and $D_{2t}$ no longer
intersect, but each intersects $E_{st}$ in a point.  In the figures
above, only $p_{41}$ and its blow up $E_{41}$ are labeled explicitly.}
\smallskip

The homology classes of the $D_{\a i}$ in the singular geometry are
\eqn\TfourSingRels{D_{1s} = \half f_1,\quad
D_{2t} = \half f_2,\quad\hbox{independent of $s$,}}
where $f_\a$ is the class of $T^2_{(\a)}$.  Let us focus on the
singularity at the ``half point''\foot{This ``half point'' is the
interpretation of $\int_{K3}(dx^1\w dx^2)\w (dx^3\w dx^4) =
\half\int_{T^4} dx^1\w dx^2\w dx^3\w dx^4 = 1/2$.} $p_{st} = D_{1s}\cap
D_{2t}$, and consider the local model $\IC^2/\IZ_2$ at this point.

\bigskip\centerline{\figscale{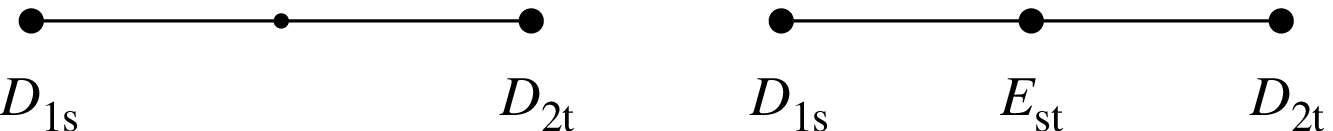}{4.0truein}}\nobreak
\caption{3}{(a) The fan for the local model $\IC^2/\IZ_2$ at the
singular point $p_{st}$ in $T^4/\IZ_2$ (left), and (b)~the fan for the
resolution (right), with the point $p_{st}$ blown up to the
exceptional divisor $E_{st}$.}
\smallskip

Fig.~3~(a) gives the fan for the toric description of $\IC^2/\IZ_2$.
There is a single two dimensional fan of volume 2 generated by the
lattice vectors $D_{1s}=(0,1)$ and $D_{2t}=(2,1)$, each of which
represents a divisor of $T^4/\IZ_2$.  If we take $p_{st}$ to be the
origin of $\IC^2/\IZ_2$, then these divisors are $D_{1s} = \{z_1=0\}$
and $D_{2t} = \{z_2=0\}$.  In the toric description, to resolve the
singularity, we subdivide the original singular cone into two cones of
volume 1 by introducing a new divisor $E_{st}$.  $E_{st}$ is the
exceptional divisor obtained by blowing up the origin of
$\IC^2/\IZ_2$.

Let us make this more explicit.  To each of the lattice components
$r$, we associate a monomial $U_r=\prod_i z_i{}^{(D_i)_r}$, where
$(D_i)_r$ is the $r$th component of the lattice vector $D_i$ in the
fan.  The toric variety is then given by the set of all $(z_1,z_2)$
not in the excluded set $F$ modulo rescalings that leave the $U_r$
invariant.  The excluded set $F$ consists of all points that have
simultaneous zeros of coordinates whose corresponding $D_i$ do not
lie in the same cone.  For the unresolved fan of Fig.~3~(a), there is
just a single two dimensional cone, so $F = \emptyset$.  The only
rescaling that leaves $U_1,U_2$ invariant is $\IZ_2\colon\
(z_1,z_2)\to(-z_1,-z_2)$.  So, the toric variety is indeed
$\{(z_1,z_2)\}/\IZ_2 = \IC^2/\IZ_2$.

For the resolved fan, we include the lattice vector $E_{st}=(1,1)$ as
well, as shown in Fig.~3~(b).  In this case, $U_1 = z_2{}^2 w$ and
$U_2 = z_1z_2w$, where $w$ is the new coordinate associated to
$E_{st}$.  The excluded set is $F=\{z_1=z_2=0\}$.  The rescaling
symmetry of $U_1,U_2$ is $\IC^*\colon\ (z_1,z_2,w)\to(\l z_1,\l
z_2,\l^{-2}w)$.  Away from $w=0$, this gives
$(z_1,z_2,1)/\IZ_2=\IC^2/\IZ_2$ with the $z$-origin deleted.  At
$w=0$, we obtain the exceptional $\IP^1$, $E_{st} =
\{(z_1,z_2,0)\,\backslash\, (0,0,0)\}/\IC^*$.

Divisors can always be represented in patches as the vanishing loci of
local meromorphic functions.  However, divisors that globally have
such a representation are homologically trivial and have trivial
intersection with other divisors.  (See, for example,
Ref.~\Griffiths.)

In our toric model for the resolution of $\IC^2/\IZ_2$, a
basis of such global meromorphic functions is $U_1,U_2$.  The
corresponding homologically trivial divisors are $2D_{1s}+E_{st}$
(from $U_2{}^2/U_1 = 0$) and $2D_{2t}+E_{st}$ (from $U_1= 0$).  In the
compact $K3$, (as explained for $\TZZ$ in Ref.~\DenefMM), these
relations become
\eqn\fKthree{\eqalign{f_1 &= 2D_{1s}+\sum_{t=1}^4 E_{st}
\quad\hbox{independent of $s$,}\cr
f_2 &= 2D_{2t}+\sum_{s=1}^4 E_{st}
\quad\hbox{independent of $t$,}}}
where the divisors $f_1$ and $f_2$ are not homologically trivial, but
instead correspond to ``sliding divisors'' that can be moved away from
the (resolved) singularities.  They have trivial intersection with the
exceptional divisors $E_{st}$ and represent the tori
$f_1=\{z_1=c_1\}\cup\{z_1=-c_1\}$ and
$f_2=\{z_2=c_2\}\cup\{z_2=-c_2\}$ on the $T^4$ covering space, where
$c_1,c_2$ are non fixed points.  The corresponding Poincar\'e dual
cohomology classes are $2dx^1\w dx^2$ and $2dx^3\w dx^4$,
respectively.

The cycles in $K3$ described so far are those that are particularly
simple in the complex structure $J_{(3)}$.  In the same way, in the
complex structure $J_{(1)}$ we obtain homology classes $f_3$ and $f_4$
from elliptic curves $T^2_{(3)}$ and $T^2_{(4)}$ located at non fixed
points in $(x^1,x^4)$ and $(x^2,x^3)$, respectively.  In the complex
structure $J_{(2)}$ we obtain homology classes $f_5$ and $f_6$ from
elliptic curves $T^2_{(5)}$ and $T^2_{(6)}$ located at non fixed
points in $(x^2,x^4)$ and $(x^3,x^1)$.  Likewise, we obtain divisors
$D_{3s},D_{4t}$ and $D_{5s},D_{6t}$ by setting the corresponding pairs
of coordinates equal to their $\IZ_2$ fixed values before the
resolution.  The homology lattice of $K3$ is the integer span of the
overcomplete basis given by the $6 f$, $24$ $D$ and 16 $E$ divisors.

To make the last paragraph more explicit, it is convenient to use the
notation of Ref.~\WendlandRY.  Let us label the fixed points by twice
their $T^4$ coordinate values (i.e, by ordered quadruples
$(y^1,y^2,y^3,y^4)$, with $y^i=2x^i\in\IF_2 = \{0,1\}$.). Then the
fixed points and exceptional divisors are specified by a point $y$ in
$\IF_2{}^4$.  The divisors $D_\k$ are
\eqn\Dkappa{D_\k = \half\k - \half\sum_{y\in P(\k)}E_y,}
where $\k$ is the pullback to $K3$ of a fixed $T^2$ in $T^4$ meeting
the subset $P\subset\IF_2{}^4$ of fixed points.  There are
$6\times4=24$ different $\k$ in $6$ homology classes $f$, from the
${4\choose2}=6$ directions spanned by the $T^2$ in $T^4$ and the
$2^2=4$ fixed locations on the transverse $T^2$.

\nonnumberedsubsec{The lattice of $\TZZ$}

The homology lattice of the Calabi-Yau orientifold $X=\TZZ$ is very
similar to that just discussed for $K3$, and is in some sense simpler
due to the absence of the hyperK\"ahler structure.  In this section,
we closely follow Ref.~\DenefMM.

The unresolved orbifold $\TZZ$ has $2^6=64$ fixed points of local
geometry $\IC^3/(\IZ_2\times\IZ_2)$.  The fan for the latter is shown
in Fig.~4~(a).  There is a single cone of volume 4, with vertices
$D_{1s}=(2,0,1)$, $D_{2t}=(0,2,1)$ and $D_{3u}=(0,0,1)$.  Here, the
second subscript of the divisor $D_{\a s}$ take values $s=1,\ldots,4$
and indicates the location of the divisor among the four fixed points
on the transverse space $\IP^1=T^2_{(\a)}/\IZ_2$.  Since we have
already reviewed the toric geometry of $T^4/\IZ_2$ above, we will be
more telegraphic here.  The monomials $U_r$ are $U_1=z_1{}^2$,
$U_2=z_2{}^2$ and $U_3=z_1z_2z_3$, with rescaling symmetry
$\IZ_2\times\IZ_2$, which we take to be generated by $\s_3$ and $\s_1$
of Eq.~\Ztwos.  The excluded set is $F=\emptyset$.  So, we indeed
obtain $\{(z_1,z_2,z_3)\}/(\IZ_2\times\IZ_2) =
\IC^3/(\IZ_2\times\IZ_2)$.

\bigskip\centerline{\figscale{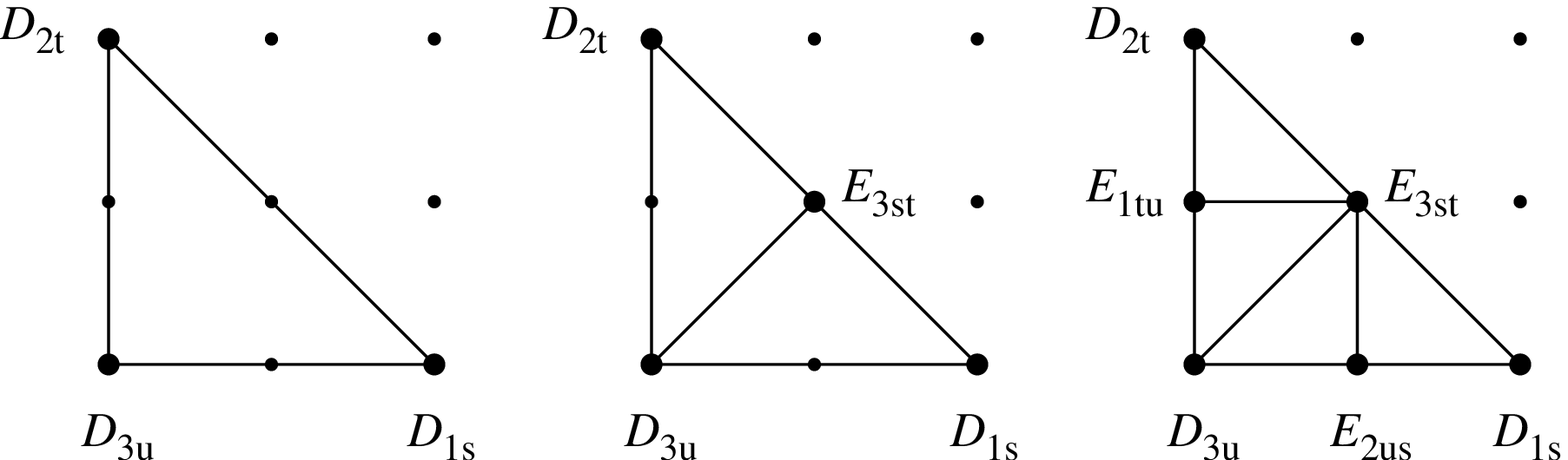}{5.0truein}}\nobreak
\caption{4}{Fans for: (a) the local model
$\IC^3/(\IZ_2\times\IZ_2)$ at the singular point $p_{stu}$ of $\TZZ$,
(b)~the partially resolved local model
$\bigl((\hbox{Eguchi-Hanson})\times \IC\bigr)/\IZ_2$, and (c) the
fully resolved local model.}
\smallskip

For the construction described in this paper we resolve $X$ one
$\IZ_2$ at a time---first with respect to $\s_3$ to obtain the
Voisin-Borcea orbifold $(K3_{(3)}\times T^2_{(3)})/\s_1$, and then
with respect to $\s_1$.  Let us focus on the local model
$\IC^3/(\IZ_2\times\IZ_2)$ at the ``quarter point''\foot{This
``quarter point'' is the interpretation of $\int_X(dx^1\w dx^2)\w
(dx^3\w dx^4)\w (dx^5\w dx^6) =\quarter\int_{T^6} dx^1\w dx^2\w dx^3\w
dx^4\w dx^5\w dx^6 = 1/4$.}  $p_{stu} = D_{1s}\cap D_{2u}\cap D_{3u}$.
The first step, the resolution with respect to $\s_3$, introduces the
exceptional divisor $E_{3st}$ in the partially resolved fan of
Fig.~4~(b).  After this step, there two cones, each of volume 2.  The
geometry is $\bigl((\hbox{Eguchi-Hanson})\times \IC\bigr)/\IZ_2$.  The
second step introduces additional exceptional divisors $E_{1tu}$ and
$E_{2us}$ and gives the fully resolved fan shown in Fig.~4~(c).  As
explained in Ref.~\DenefMM, this is the {\it asymmetric\/} resolution
of $\IC^3/(\IZ_2\times\IZ_2)$ with distinguished direction $\a=3$.  A
{\it symmetric\/} resolution is also possible, and differs by a flop
of the curve $C_{3stu} = D_{3u}\cap E_{3st}$ (the diagonal line in
Figs.~4~(b,c)).

Let $w_1,w_2,w_3$ denote the toric coordinates associated to
$E_{1tu},E_{2us},E_{3st}$, respectively.  Then, in the resolution of
the local model, we obtain the monomials $U_1=z_1{}^2w_2w_3$,
$U_2=z_2{}^2w_3w_1$ and $U_3=z_1z_2z_3w_1w_2w_3$.  The scaling
symmetry is
\eqn\TZZrescaling{{\IC^*}^3\colon\quad(z_1,z_2,z_3,w_1,w_2,w_3)
\to(\l_1z_1,\l_2z_2,\l_3z_3,{\l_1\over\l_2\l_3}w_1,
{\l_2\over\l_3\l_1}w_2,{\l_3\over\l_1\l_2}w_3),}
and the resolved local model is the toric variety
$(\IC^6\,\backslash\,F)/{\IC^*}^3$, where $F$ is the excluded set.

From fact that $U_1,U_2,U_3$ are meromorphic functions in the local
model, we obtain the homology relations $2D_{1s}+E_{2us}+E_{3st}=0$ in
the local model, together with cyclic permutations in $1,2,3$.  In the
compact Calabi-Yau manifold $X$, these relations become
\eqn\TZZrelations{\eqalign{%
F_1 &= 2D_{1s} + \sum_{u=1}^4 E_{2us}+\sum_{t=1}^4 E_{3st}
\quad\hbox{independent of s,}\cr
F_2 &= 2D_{2t} + \sum_{s=1}^4 E_{3st}+\sum_{u=1}^4 E_{1tu}
\quad\hbox{independent of t,}\cr
F_3 &= 2D_{3u} + \sum_{t=1}^4 E_{1tu}+\sum_{s=1}^4 E_{2us}
\quad\hbox{independent of u.}}}
Here, the $F_\a$ are homologically nontrivial ``sliding divisors,''
which can be moved away from the (resolved) singularities of $X$, and
therefore do not intersect the exceptional cycles.  These are the
divisors $K3_{(\a)}$ already mentioned in Sec.~3.1.

The integer homology lattice $H^2(X,\IZ)$ is generated by
$\{F_\a,D_{\a s},E_{\a st}\}$.  Using the relations \TZZrelations, the
subset $\{F_\a,E_{\a st}\}$ forms a linearly independent basis, though
one that requires some coefficients in $\IZ/2$ rather than $\IZ$ to
form a basis for $H^2(X,\IZ)$.  (For example, $D_{1s} = \half F_1 -
\half \sum_{u=1}^4 E_{2us}-\half\sum_{t=1}^4 E_{3st}$.)  This is the basis
employed throughout the paper, with the following modification: for
notational simplicity we use the multi-index $I$ instead of $st$ for
the exceptional divisors.

The intersection numbers $\k_{abc}$ for $a,b,c$ distinct can be
computed using the local model.  The remaining intersections,
$\k_{aab}$ and $\k_{aaa}$ then follow from the $\k_{abc}$ together
with the relations \TZZrelations, as discussed in Ref.~\DenefMM.  Let
us simply quote the result here.  If we write the Poincar\'e dual of
the K\"ahler form as
\eqn\XKahler{J = v^iF_i - v^{3st}E_{3st} - v^{1tu}E_{1tu}
- v^{2us}E_{2us},}
then the volume of $X$ is
\eqn\TZZintersections{\eqalign{V_X 
&= {1\over6}\k_{abc}v^av^bv^c = 2v^1v^2v^3 
-v^3\sum_{st}(v^{3st})^2 - v^1\sum_{tu}(v^{1tu})^2
-v_2\sum_{us}(v^{2us})^2\cr
&\qquad 
-{4\over3}\Bigl(\sum_{tu}(v^{1tu})^3+\sum_{us}(v^{2us})^3\Bigr)
+\sum_{stu}v^{3st}\Bigl((v^{1tu})^2+(v^{2us})^2\Bigr).}}
This is twice the volume given in Eq.~(6.9) of Ref.~\DenefMM.  Note
that for $v^{1tu}=v^{2us}=0$, which is the case for the partial
resolution of Fig.~4(b), we obtain half of the intersection form of
$K3^{\vphantom2}_{(3)}\times T^2_{(3)}$ expressed as a function of the
$T^2_{(3)}$ K\"ahler modulus $2v^3$ and $K3_{(3)}$ K\"ahler moduli
$v^a$, $a\ne3$.  Likewise for $a=F_\a$, the $\k_{abc}$ give the
correct intersection numbers $\k_{bc}$ on $K3_{(\a)}$.


\appendix{C}{Calabi-Yau manifolds of Voisin-Borcea type}

Calabi-Yau manifolds of the form $X=(K3\times T^2)/\IZ_2$ comprise the
Voisin-Borcea class \refs{\BorceaTQ,\Voisin}.  They are conventionally
characterized by three integers $(r,a,\d)$, which we now describe.

The $\IZ_2$ acts by inversion ($\CI_2:\ x^n\to-x^n$) on the $T^2$ and
holomorphic involution $\s$ on the $K3$ surface.  In order that there
exist a $(3,0)$ form on the quotient $X$, the involution must act as
$(-1)$ on $H^{2,0}(K3)$.  Nikulin \Nikulin\ has classified all such
involutions in terms of the integers $(r,a,\d)$.

The integer $r$ was already discussed in Sec.~5.  The second
cohomology lattice of $K3$ is
\eqn\Htwolattice{H^2(K3,\IZ) = (- E_8\times E_8)\times U_{1,1}^3,}
where $U_{1,1}$ is a two dimensional lattice of signature $(1,1)$.
Under the action of $\s$, $H^2(K3,\IZ)$ decomposes into even and odd
parts $H^2_+(K3,\IZ)$ and $H^2_-(K3,\IZ)$.  The integer $r$ gives the
rank of $H^2_+(K3,\IZ)$.  In terms of $r$, the orbifold untwisted
sector of $X$ contributes
\def\ut{{\rm ut}}
\def\tw{{\rm tw}}
\eqn\utVB{h_\ut^{1,1}=r+1\quad\hbox{and}\quad h_\ut^{2,1}=21-r.}
For $\TZZ$ we have $r=18$, so $(h_\ut^{1,1},h_\ut^{2,1})=(19,3)$ at
this stage, after resolution of a single $\IZ_2$.

The integer $a$ determines the Hodge numbers in the orbifold twisted
sector. In this sector, the blow-ups and complex structure
deformations of the orbifold singularities contribute
\eqn\twVB{\eqalign{h_\tw^{1,1} &= 4+2r-2a = 4(k+1),\cr
h_\tw^{2,1} &= 44-2r-2a = 4g.}}
Let us focus on the rightmost expressions.  The factors of 4 arise
since there are four fixed points on the base $\IP^1 = T^2/\IZ_2$.
Over each fixed point there are $k+1$ fixed curves: $k$ rational
curves and a single genus $g$ curve.  Each of the $k+1$ curves
contributes one K\"ahler modulus and the genus $g$ curve contributes
$g$ complex structure deformations.  It can be shown that once $r$ is
specified, $k$ and $g$ are not independent.  They can be parametrized
in terms of a single integer $a$, with $r-a$ even, as $k=\half(r-a)$
and $g=\half(22-r-a)$.\foot{There are two exceptions to the statements
in this paragraph: When $(r,a,\d) = (10,10,0)$, the involution $\s$
acts freely on $K3$ to give an Enriques surface with no fixed points.
This is the FHSV case \FerraraYX.  For $(r,a,\d)=(10,8,0)$, we obtain
the disjoint union of two elliptic curves instead of one rational
curve and one elliptic curve.}

For $\TZZ$ this gives 8 fixed $\IP^1$s over each fixed point.  In
terms of the $T^4/(\IZ_2\times\IZ_2)=\IP^1_{(1)}\times\IP^1_{(2)}$
fiber, these can be interpreted as
$$(4\hbox{ fixed points on }\IP^1_{(1)})\times\IP^1
\quad\cup\quad
\IP^1\times(4\hbox{ fixed points on }\IP^1_{(2)}).$$

From Eq.~\twVB, we see that the subset of Voisin-Borcea manifolds that,
like $\TZZ$, have $h_\tw^{2,1}=0$ (and hence have the same spectrum of
3-cycles on which D6 branes can be wrapped) are of the form
$(r,a,\d)=(r,22-r,\d)$.  Since $r\ge a$, this means that $r\ge11$.
Interestingly, the condition $r+a=22$ corresponds to the cases that
the lattice $H^2(K3,\IZ)$ cannot be primitively embedded in $(-
E_8\times E_8)\times U_{1,1}^2$ \BorceaTQ.\foot{There is one other
exceptional case, $(r,a,\d) = (14,6,0)$, for which the lattice
$H^2(K3,\IZ)$ cannot be primitively embedded in $(- E_8\times
E_8)\times U_{1,1}^2$.}  When such cases are excluded, the possible
triples $(r,a,\d)$ are symmetric about $r=10$.

One additional property to note is the Euler characteristic,
\eqn\VBeuler{e = 2(h^{1,1} - h^{2,1}) = 12(r-10).}
Independent of the particular realization of the geometric twists
described in Sec.~2, we see that elements of $H^{1,1}(X)$ and
$H^{2,1}(X)$ are lifted in pairs, one from each group. Thus,
$|e/12| = |r-10|$ sets a lower bound on the number of moduli that
remain unlifted by the geometric twists alone.  (Of course the NS and
RR fluxes can lift additional moduli).

The final integer $\d$ takes values 0 or 1.  We set $\d=0$ if the
fixed locus of $\s$ on $K3$ is a class divisible by 2 in $H^2(K3,\IZ)$
and $\d=1$ otherwise.  When $r\equiv 2\ ({\rm mod}\ 4)$, both values
of $\d$ are possible and in most cases both occur, but for other $r$
only $\d=1$ is possible.  We refer the reader to
Refs.~\refs{\BorceaTQ,\Nikulin} for a table of all possible
$(r,a,\d)$.

\listrefs
\bye